\documentclass[twocolumn,preprintnumbers,nofootinbib]{revtex4}

\usepackage{graphicx}
\usepackage{amsmath,amssymb}
\usepackage{mathrsfs}
\usepackage{siunitx}
\usepackage{url}
\usepackage{physics}
\usepackage{tensor}
\usepackage{placeins}

\usepackage{subfigure}
\usepackage{natbib}
\usepackage{array, booktabs} 

\usepackage{xcolor}
\definecolor{darkBlue}{rgb}{0, 0, 0.8}

\usepackage{hyperref}
\hypersetup{
  bookmarksopen=true,     
  colorlinks=true,        
  linkcolor=darkBlue,     
  citecolor=darkBlue,      
  filecolor=darkBlue,      
  urlcolor=darkBlue        
}

\def\i{{\mathrm i}}
\def\d{{\mathrm d}}

\def\e{{\mathrm e}}

\def\L{{\mathscr L}}

\def\theta{\vartheta}
\def\eps{\varepsilon}

\def\und{{\qquad\mbox{and}\qquad}}

\renewcommand{\vec}[1]{\boldsymbol{#1}}
\def\be{\begin{equation}}
\def\ee{\end{equation}}
\def\ba{\begin{eqnarray}}
\def\ea{\end{eqnarray}}
\def\nn{\nonumber}

\def\lsim{\raise0.3ex\hbox{$\;<$\kern-0.75em\raise-1.1ex\hbox{$ m\;$}}}
\def\gsim{\raise0.3ex\hbox{$\;>$\kern-0.75em\raise-1.1ex\hbox{$ m\;$}}}

\begin{document}

\title{Lunar response to gravitational waves}

\author{M.~Kachelrie\ss}
\author{M.~P.~N\o dtvedt}

\affiliation{Institutt for fysikk, NTNU, Trondheim, Norway}


\begin{abstract}
It has been suggested to use seismic detectors on the Moon as a tool
to search for gravitational waves in an intermediate frequency range
between mHz and Hz. Employing three different spherically symmetric
models for the lunar interior, we investigate the response of the Moon
to gravitational waves in Einstein and Jordan-Brans-Dicke gravity. We
find that the first eigenfrequencies of the different models depend only
weakly on the model details, with the fundamental frequency $\nu_1$
close to 1\,ms both for spheroidal and toroidal oscillations. In contrast,
the resulting displacement varies up to a factor five, being in the
range (3.6 $\times10^{12}$--1.9$\times 10^{13})/h_0$\,cm
for spheroidal oscillations with amplitude
$h_0$ and assuming a quality factor $Q_n=3300$. Toroidal
oscillations are suppressed by a factor \(2\pi\nu R/c\),
both in Einstein gravity and in general scalar-tensor theories.
\end{abstract}


\maketitle


\section{Introduction}

Historically, the excitation of vibrational eigenmodes in an elastic body was
one of the first signatures suggested as proof for the existence of
gravitational waves (GW). In addition to the use of resonant bars on laboratory
scale, Weber also pointed out that GWs could be searched for monitoring the
vibrations of the Earth or Moon~\cite{WeberPhysRev.117.306}. The first
calculation of the response of the Earth to a GW was performed soon later
by Dyson, assuming a flat homogeneous Earth model~\cite{Dyson:1969zgf}.
The response of the Earth to a GW for a spherically, heterogeneous Earth model
was first determined by Ben-Menahem~\cite{Ben-Menahem:1983}, who followed the
approach developed by Alterman {\it et al.}~\cite{Alterman59} for the study
of seismic waves in the Earth.

Searches for GW using seismographs on Earth started already in the
1970s~\cite{https://doi.org/10.1029/JB074i022p05351,1974ApJ...187L..49M}.
More recently, seismic data were used to derive stringent limits on the
stochastic GW background in the frequency range
0.05--1\,Hz~\cite{Coughlin:2014xua}. In Ref.~\cite{Majstorovic:2019fog},
the response of a nonrotating anelastic Earth model to a GW was
revisited. On the Moon, the Lunar Surface Gravimeter
experiment was deployed by Apollo~17, but technical problems prevented the
usage of its data. In the last years, the idea to use the Moon as
GW detector has been revived and several new concepts were proposed:
One type of experiments proposes to construct long-baseline interferometers
similar to the successful LIGO set-up, as e.g.\ the
LION proposal~\cite{Amaro-Seoane:2020ahu} or the Gravitational-Wave Lunar
Observatory for Cosmology GLOC~\cite{Jani:2020gnz}. Another type of proposal
aims to exploit the response of the Moon to GWs similar to the original
Weber suggestion, as e.g.\ the Lunar Gravitational-Wave
Antenna (LGWA) experiment~\cite{LGWA:2020mma} or the 
Lunar Seismic and Gravitational-Wave Antenna (LSGWA)~\cite{LSGA}.
These lunar GW experiments could become an important partner observatory
for joint observations with the space-borne, laser-interferometric detector
LISA~\cite{LISA:2017pwj} and the planned underground Einstein
observatory~\cite{ET}, exploiting the
weak seismic activity of the Moon~\cite{ALSEP}. In particular, they
could complement these observatories in the mHz range where their
sensitivity has been estimated to be superior~\cite{LGWA:2020mma}.
For instance, GWs from binary white-dwarf systems  could be searched for
by  matching the frequencies of Moon's normal-modes  with the wave-forms
expected for these binaries~\cite{Kupfer:2018jee}. For such searches,
a precise understanding of the response of the Moon to GWs is a
pre-requisite.

In this work, we study the response of the Moon to gravitational
perturbations employing and extending the approach of
Ref.~\cite{Ben-Menahem:1983}. We derive a set of first-order differential
equations which determine the eigenfunctions and eigenfrequencies of the
Moon coupled to a GW for a given  spherically, heterogeneous Moon model.
We account for a potentially scalar polarization state in the GW, so
that our results are also valid for general scalar-tensor theories of
gravity like, e.g.,  Jordan-Brans-Dicke theories~\cite{Jordan59,Brans:1961sx}.
We determine the displacement and the eigenfrequencies of the first
eigenmodes numerically for a set of three different Moon models.
We find that there is very good agreement on the eigenfrequencies in all
three models, while the magnitude of the displacement varies up to factor
two. Using the predicted capability to measure ground displacement in the
LGWA experiment from Ref.~\cite{van_Heijningen_2023}, we find a nominal
sensitivity to GWs with amplitude  $h\simeq 10^{-20}$ in the mHz
range assuming as quality factors $Q_n\simeq 3300$.

This work is structured as follows: In Section~II, we recall the response of
an elastic body to a GW in a general metric theory of gravity. We derive in
Section~III the normal modes of the Moon, and summarize how its
eigenfrequencies and displacements can be numerically calculated;
most technical details of this derivation are deferred into two appendices.
Section~IV
introduces the models used to describe the Moon and presents our numerical
results. Finally, we make concluding remarks in Section~V.

\section{Response of an elastic body to a GW}

An elastic isotropic body with density $\rho$ can be described in the
non-relativistic limit by the Lagrange density
\be
 \L=\frac12 \rho \dot u_i\dot u^i -\frac12\eps_{ij} \sigma^{ij} ,
\ee
where $u^i$ denotes the displacement of a body element.
Its strain tensor $\eps_{ij}$ and stress tensor $\sigma_{ij}$ are connected
by the Cauchy relation,
\be
 \sigma_{ij} = \lambda\delta_{ij} \nabla_k u_{k}
 +\mu( \nabla_i u_{j}+ \nabla_j u_{i}) =
\lambda\delta_{ij} \nabla_k u_{k} + 2\mu \eps_{ij} .
\ee
Both tensors depend on the two Lam\'e parameters $\lambda$ and
$\mu$ which determine the response of the body to  bulk and shear forces.

The coupling of matter to an external gravitational perturbation
$h_{\mu\nu}$ is at lowest order perturbation theory given by
$\L_{\rm int}= \kappa T_{\mu\nu} h^{\mu\nu}$ with $ T_{\mu\nu}$ as the
(relativistic) stress energy-momentum tensor and $\kappa=8\pi G_N$
as the gravitational coupling.  In the following, we do not impose the
transverse-traceless (TT) gauge condition on $h^{\mu\nu}$. Instead, we assume
only that gravitational waves (GW) satisfy $h_{0\mu}=h_{\mu 0}=0$.
Thus we allow in particular for the possible presence of a scalar 
polarization state which might arise in theories of modified
gravity. Then the Lagrange density describing the elastic body under the
influence of a GW is given by
\be
\L=\L_0+\L_{\rm int} =
\frac12 \rho \dot u_i\dot u^i -\frac12(\eps_{ij}+h_{ij}) \sigma^{ij} .
\ee
Thus the GW acts, as expected,  as an additional strain on
the body. The equation of motion of the body follows as
\be      \label{EoMgenral}
\partial_t(\rho u^i) = \nabla_j \sigma^{ij} - \nabla_j(\mu h^{ij})
- \frac12 \nabla^i( \lambda h) .
\ee
The last two terms represent the driving force density \(f_i\)
exerted by the GW on the body,
\begin{equation}
    f_i = - \nabla_j(\mu h_i^{\;j})
- \frac12 \nabla_i( \lambda h).\label{eqn:GW_driving_force}
\end{equation}
In Einstein gravity, where $h\equiv h_i^{\;i}=0$  is valid in a physical
gauge, the last term is absent. The corresponding stress is given by
\begin{equation}
  \sigma_{ij} = -\mu h_{ij} - \frac{1}{2}\lambda h\delta_{ij} .
  \label{eqn:GW_surface_stress}
\end{equation}

The GW in Eq.~(\ref{EoMgenral}) can be represented as a superposition of
monochromatic polarization states,
\be
 h_{ij}(t)= g(t) h_0  \mathcal{E}_{ij}  \sin(\omega t) ,
 \ee
with amplitude $h_0$ and a time-dependent modulation given by
$0\leq g(t)\leq 1$.
The polarization tensor $\mathcal{E}_{ij}$ contains in a general
scalar-vector-tensor theory of gravity six independent components,
$A_S, A_L, A_{V_1}, A_{V_2}, A_+$ and $A_\times$, see for an extended discussion
e.g.~Ref.~\cite{Poisson14}. If the GW travels in the $z$
direction, the polarization tensor has the form
\begin{equation}
\mathcal{E}_{ij} = 
\begin{bmatrix}
A_S + A_+ & A_\times & A_{V_1}\\
A_\times & A_S - A_+ & A_{V_2}\\
A_{V_1} & A_{V_2} & A_L
\end{bmatrix} .
\label{eqn:chapter3_polarisationtensor}
\end{equation}
In addition to the two polarization states $A_+$ and $A_\times$ present in
Einstein gravity, two transverse $(A_{V_1}, A_{V_2})$ and one longitudinal
$A_L$ vector component, as well as the scalar component $A_s$
may enter $\mathcal{E}_{ij}$. As scalar extensions of Einstein gravity
are far more popular than vector ones, we will neglect for simplicity
the vector components in the following.

\section{Normal modes of a spherical self-gravitating body}

\subsection{Normal modes of a spherical body}

The perturbations of a spherically symmetric body factorize in the
variables $t,r,$ and $\theta,\phi$.  They are characterized by a set of
eigenfunctions and eigenfrequencies which are specified by the three
``quantum numbers''  $\{n,m,l\}$.  In a spherically symmetric body, the
modes are degenerate in $m$. Neglecting for the moment the
time-dependence, the displacement vector for a given mode $\{n,m,l\}$
can be written as a linear combination of the three Hansen vectors, or
equivalently as a sum of the vector
surface harmonics $\vec C^{(ml)}, \vec P^{(ml)}$, and $\vec B^{(ml)}$,
 \begin{align}
 \vec u(\vec{r}) &= U^{(n)}(r) \vec P^{(ml)}(\theta,\phi)
   + V^{(n)}(r)\sqrt{l(l+1)}\vec B^{(ml)}(\theta,\phi)\nn\\ 
   & + W^{(n)}(r)\sqrt{l(l+1)}\vec C^{(ml)}(\theta,\phi).
   \label{eqn:displ_gen}
\end{align}
In spherical coordinates, the vector surface harmonics are given by
\begin{subequations}
    \begin{equation}
        \sqrt{l(l+1)}C^{(ml)}_i(\theta,\phi) = \left(\hat{e}^{(\theta)}_i\frac{1}{\sin{\theta}}\frac{\partial}{\partial\phi}-\hat{e}^{(\phi)}_i\frac{\partial}{\partial\theta}\right)Y^{(ml)},
    \end{equation}
    \begin{equation}
        P_i^{(ml)}(\theta, \phi) = \hat{e}^{(r)}_iY^{(ml)}(\theta,\phi),
    \end{equation}
    \begin{equation}
        \sqrt{l(l+1)}B^{(ml)}_i(\theta, \phi) =\eps_{ijk} \hat{e}^{(r)}_j C_k^{(ml)},
    \end{equation}
\end{subequations}
where $\hat{\vec e}^{(j)}=\{\hat{\vec e}^{(r)},\hat{\vec e}^{(\phi)},
\hat{\vec e}^{(\theta)}\}$ are orthonormal unit base vectors,
$\eps_{ijk}$ denotes the Levi-Civita symbol,
\(P^{(ml)}(\theta,\phi)\) is the associated Legendre polynomial, and the
spherical harmonics \(Y^{(ml)}(\theta,\phi)\) are defined in
Eq.~(\ref{spherHarm}).
In order to distinguish the indices labeling eigenmodes from
coordinate indices, we set the former in parentheses. 
Since we use unit base vectors, we do not need to distinguish between upper
and lower indices.

The eigenmodes can be split into two independent sets of modes:
spheroidal oscillations (with $W^{(n)}=0$) which modify the shape of the
body and toroidal oscillations (with $U^{(n)}=V^{(n)}=0$) which do not.


The time dependence of small oscillations
$\vec u(\vec{r},t)=\vec u(\vec{r})\bar g(t)$ of an
elastic body are naturally modeled as a damped
harmonic oscillator. Thus the time-dependent  effect of a monochromatic GW
with frequency $\omega_0$ on a elastic body follows as a Fourier integral
of the Green function $ G_n(\omega)$ of a forced damped harmonic oscillator
with eigenfrequency $\omega_n$ weighted by $g(\omega)$, 
\begin{subequations}
\begin{align}
\bar g_n(t) & =\int \frac{{\rm d}\omega}{2\pi} \:
g(\omega)\e^{\i\omega_0 t} G_n(\omega)
  \\ & =
\int \frac{{\rm d}\omega}{2\pi} \:
\frac{g(\omega)\e^{\i\omega_0 t}}{\omega_n^2-\omega^2 +\i\omega_n\omega/Q_n} .
\label{g_bar}
\end{align}
\end{subequations}
Here, $g(\omega)$ is the Fourier transform of $g(t)$, while $Q_n$ is the
damping (or quality) factor of the mode $n$.
Note that other choices for the Green function are in use, which
should differ mainly in how anharmonic terms in the response to the GW are
parametrised. 
Our default choice of a forced damped harmonic oscillator is the one often
employed in seismology, but we will present later results also for another
Green function.

As a simple model for $g(t)$, we consider for illustration a finite
monochromatic GW of duration $2\tau$,
\begin{equation}
    g(t) = [\theta(t+\tau)-\theta(t-\tau)] \e^{\i\omega_0 t},
\end{equation}
with $\theta(t)$ as the Heaviside step function.
Then $\bar g_n(t)$ follows as 
\begin{equation}
    \bar g_n(t)  \simeq \frac{g(t)}{(\omega_0-\omega_n)^2+\omega_n^2/(4Q_n)}
\label{barg}
  \end{equation}
in the limit $Q_n\gg 1$.

\subsection{Linearisation of Euler and Poisson equations}

In Fourier space, the Euler equation becomes in spherical coordinates
\begin{equation}
 \frac{\partial}{\partial r}\sigma_{ir} + \frac{1}{r}\frac{\partial}{\partial\theta}\sigma_{i\theta}  + \frac{1}{r\sin{\theta}}\frac{\partial}{\partial\phi}\sigma_{i\phi}+ \rho F_i + \rho\omega^2u_i = 0.
    \label{eqn:CauchysEquation}
\end{equation}
We simplify this equation using the following assumptions:
First of all, we restrict ourselves to linear perturbations. Then we
assume that the self-gravitating body is initially in equilibrium
between the hydrostatic pressure gradient $\vec\nabla P^{(0)}$ and internal
gravitational forces,  $\vec F^{(0)}=\vec f^{(0)}/\rho=\vec\nabla\Psi^{(0)}$.
Here, we introduced the gravitational (anti-) potential $\Psi^{(0)}$ and 
we denote unperturbed quantities with the subscript zero.
Being strained, a volume element carries its initial stress
$\sigma^{(0)}(\vec r^{(0)})$ to its new position $\vec r^{(0)} +\vec u$.
Thus
\begin{subequations}
\begin{align}
  \sigma_{ij}^{(0)}(\vec r^{(0)}) &=  \sigma^{(0)}_{ij}(\vec r -\vec u) =
  -P^{(0)} (\vec r -\vec u) \delta_{ij}(\vec r) 
   \\ &
  = -P^{(0)} (\vec r)
  -\vec u(\vec r) \vec\nabla P^{(0)}(\vec r) \delta_{ij}(\vec r)\\ 
  &= -(P^{(0)} + g^{(0)}\rho^{(0)} u_r) \delta_{ij}(\vec r) ,
\end{align}
\end{subequations}
where $g^{(0)}$ is the unperturbed gravitational acceleration.
In addition, the volume element aquires an
additional stress $\delta\sigma_{ij}$ after displacement due to
distortions, given by the usual Cauchy relation. Thus
\begin{align}
  \delta\sigma_{ij} &= 
\lambda\delta_{ij} \nabla_k u_{k} + 2\mu \eps_{ij} .
\end{align}
We can now insert the total stress
\(\sigma_{ij} = \sigma_{ij}^{(0)}+\delta\sigma_{ij}\) into
Eq.~(\ref{eqn:CauchysEquation}). The divergence of the
initial stress becomes
\begin{align}
    \nabla_k(\sigma_{ik}^{(0)}) &=-\nabla_k(P^{(0)}+g^{(0)}\rho^{(0)}u_r)\\
    &=  g^{(0)}\rho^{(0)}\hat{e}_k^r-\frac{\d\rho^{(0)}}{\d r}g^{(0)}u_r-\rho^0\nabla_k(g^{0}u_r),\nn
\end{align}
while the force term can be written as
\begin{align}
    \rho F_k &= \left(\rho^{(0)}-\frac{\d\rho^{0}}{\d r}u_r - \rho^{(0)}\nabla_iu_i\right)(\nabla_k\psi-g^{(0)}\hat{e}_k^r)\\
    &= \rho^{(0)}\nabla_k\psi + g^{(0)}\left(\frac{\d\rho^{(0)}}{\d r}u_r+\rho^{(0)}\nabla_ku_k-\rho^{(0)}\right)\hat{e}_j^{r},\nn
\end{align}
where the continuity equation $
\rho-\rho^{(0)}=\vec\nabla\cdot (\rho^{(0)}\vec u)$ was used in the first step
to expand $\rho$. Including the remaining part of
Eq.~(\ref{eqn:CauchysEquation}), we arrive
at the following differential equation
\begin{align}
  &  \nabla_k(\lambda\nabla_i u_i) + \mu[\Delta u_k + \nabla_k(\nabla_iu_i)]
 \nn   \\ &+ \frac{\d\mu}{\d r}\left(2\frac{\partial u_k}{\partial r} + [\hat{e}^{(r)}\times(\nabla\times \mathbf{u})]_k\right)    \label{euler}
 \\ & + \rho^{(0)}\nabla_k(\psi-g^{(0)}u_r) + \rho^{(0)}\hat{e}^{(r)}_k\nabla_iu_i + \omega^2\rho^{(0)}u_k = 0 . \nn
\end{align}
In addition, the potential $\Psi = \Psi^{(0)} + \psi$
obeys the Poisson equation, implying for the perturbation
\begin{equation}
  \Delta\psi = -4\pi G\left(\rho-\rho^{(0)}\right)
  = 4\pi G\nabla_j\left(\rho^{(0)} u_j\right).
    \label{poisson}
\end{equation}
Equations~\eqref{euler} and \eqref{poisson} are the two coupled differential
equations which we have to solve
numerically under appropriate boundary conditions. We assume that the Moon,
similar to the Earth, can be divided into a liquid core and a solid mantle.
In this case, one can follow  the procedure developed for the
study of seismic waves in the Earth, as described in detail e.g.\
in Ref.~\cite{SeismicWaveAndSources}. For the convenience of the
reader, the transformation of Eqs.~\eqref{euler} and
\eqref{poisson} to a set of linear differential
equation for $\{y_1,\ldots,y_6\}$ is summarized in the
Appendix~\ref{reduction}.

\begin{widetext}

\subsection{Normal modes of a spherical self-gravitating body}

As a consequence of our linearisation, the Fourier  modes $\{nml\}$
decouple and we can consider the evolution of a single mode. Moreover,
our assumption of spherical
symmetry implies that the Euler and Poisson equations reduce to ordinary
differential equation in the radial coordinate, while the angular
dependence  can be expressed as  Fourier transforms of the vector surface
harmonics on $S^2$, which we define as
\be  \label{FourierV}
\widetilde V^{(lm)}_{ij}(kr)\equiv
\int_0^{2\pi}\d\phi\int_0^\pi\d\theta \sin{\theta}\:
\hat{e}^{(r)}_i V^{(lm)}_j \e^{-\i\vec k\vec r}
\ee
for the three cases
$\vec V^{(lm)}=\{\vec C^{(lm)},\vec P^{(lm)},\vec B^{(lm)}\}$.

The displacement of the mode $\{nml\}$ induced by the force distribution
\(f^{(n)}_i(\vec{r}^{(0)})\) and the surface stresses
\(\sigma^{(0)}_{ij}(\vec{u})\) to a radially heterogeneous, anelastic
self-gravitating Moon model can be derived from\footnote{For a text book discussion see Ref.~\cite{Ben-Menahem:1983}.} 
\begin{align}
  u^{(nml)}_i(\vec{r},t) = &\int_V \mathcal{G}^{(nml)}_{ji}(\vec{r}|\vec{r}^{(0)},t) f^{(n)}_j(\vec{r}^{(0)}) \d^3\vec{r}^{(0)}
 + \int_S \mathcal{G}^{(nml)}_{ji}(\vec{r}|\vec{r}^{(0)},t)\hat{e}^{(r,0)}_k\sigma^{(0)}_{kj}(\vec{u}) \d S(\vec{r}^{(0)})
    \label{eqn:spheroidal_displacement_1}
\end{align}
knowing the Green function \(\mathcal{G}^{(nml)}\).  The Green function
\begin{equation}
    \mathcal{G}^{(nml)}_{ij}(\vec{r}|\vec{r}^{(0)},t) = Q^{*(nlm)}_i(\vec{r})Q_j^{(nlm)}(\vec{r}^{(0)})\Bar{g}^{(n)}(t)(\Lambda_T^{(nml)})^{-1}
   \label{eqn:Greens}
\end{equation}
is in turn constructed out of the tensor product over the eigenvectors
$Q_j^{(nlm)}$
for toroidal and spheroidal oscillations, respectively, their normalization,
\begin{align}
  \Lambda_T^{(nml)} &= \frac{4\pi}{2l+1} \,l(l+1)
                      \int_0^R[y_{1n}^T]^2\rho^{(0)}(r)r^2\d r,
  \label{normT}\\
    \Lambda_S^{(nml)} &= \frac{4\pi}{2l+1}\int_0^R\left([y^S_{1n}]^2+l(l+1)[y^S_{3n}]^2\right)\rho^{(0)}(r)r^2\d r ,
\label{normS}
\end{align}
and the time-dependence given by $\bar g(t)$.
Inserting the driving force \eqref{eqn:GW_driving_force} and the surface stress \eqref{eqn:GW_surface_stress} induced by a GW, we can rewrite
Eq.~\eqref{eqn:spheroidal_displacement_1} as
\begin{align}
    u^{(nml)}_i(\vec{r},t) = &-\int_V\frac{\d\mu}{\d r}(r)\mathcal{G}^{(nml)}_{ij}(\vec{r}|\vec{r}^{(0)},t)  \hat{e}^{(r)}_k h^{kj} \,\d^3r^{(0)}
  + \mu(R)\int_S \mathcal{G}^{(nml)}_{ij}(\vec{r}|\vec{R},t)\hat{e}^{(r)}_k h^{kj} \,\d S^{(0)}(\vec{R})\nn\\
  &-\frac{1}{2}\int_V\frac{\d\lambda}{\d r}(r)\mathcal{G}^{(nml)}_{ij}(\vec{r}|\vec{r}^{(0)},t) \hat{e}^{(r)}_jh\,\d^3r^{(0)}
  + \frac{1}{2}\lambda(R)\int_S \mathcal{G}^{(nml)}_{ij}(\vec{r}|\vec{R},t) \hat{e}^{(r)}_j h \,\d S^{(0)}(\vec{R}).
\end{align}
To proceed, we have to distinguish between toroidal and spheroidal modes,
using the results for the parameter functions $y_i$ obtained in the
appendix. We will start with the simpler case of toroidal oscillations,
checking explicitly  that also the new contributions induced by the scalar
polarization state in the GW are suppressed.

\subsubsection{GW Induced toroidal motion}

In the toroidal case, we can simplify the displacement
of the mode $\{nml\}$ using Eq.~\eqref{eqn:toroidal_displacement} to
\begin{equation}
  \vec{u}^{(nml)}_j(\vec{r},t) = l(l+1) h_0\Bar{g}(t) (\Lambda^{(nml)})^{-1}
  y_{1n}(r) C^{*(lm)}_j(\theta,\phi) F_T
    \label{eqn:Toroidal_oscillation_response}
\end{equation}
with
\begin{align}
  F_T =&  R^2y_{1n}(R)\left[
       \mu(R) \mathcal{E}^{ij}+\frac12\lambda(R)\delta_{ij}\right]
     \widetilde C^{(lm)}_{ij}(kR) 
  -\int_0^R\d r \, r^2 y_{1n}(r)\left[
    \frac{\d\mu(r)}{\d r}\mathcal{E}^{ij} +
    \frac12\frac{\d\lambda(r)}{\d r} \delta_{ij} \right]
  \widetilde C^{(lm)}_{ij}(kr),
       \label{eqn:chapter4:F_T_first_expression}
\end{align}
where we introduced the Fourier transform of the vector surface harmonics
defined in Eq.~(\ref{FourierV}).

Following Ben-Menahem~\cite{Ben-Menahem:1983}, we take now advantage of
the low-frequency limit: Since the fundamental frequency $\nu_1$ of the
Moon is in the mHz range, it holds $ c/2\pi\nu_1 \gg R$. This implies that
the arguments $kr$ and  $kR$ of the exponential functions in
$\widetilde C$ are slowly varying. Performing then a partial-wave
expansion of $\widetilde C$, only the partial waves with the lowest angular
momentum have to be kept. We split the calculation into a $\mu$ and a $\lambda$ 
dependent part; the latter is absent in Einstein gravity. In the appendix,
we elucidate the details of the partial-wave expansion of  $\widetilde C$,
which results for the term including \(\mu\) and its derivative in
\begin{align}
  \frac{\sqrt{l(l+1)}}{8\pi\i}\,F_T^{(\mu)}  & =
  \mu(R)R^2y_{1n}^T(R)\sum_{l_1=0}^\infty\sum_{m_l = -l_1}^{l_1}(2l_1+1)
    \i^{-l_1}j_{l_1}(k R)Y_{l_1}^{*m_1}(0,0)
    \begin{pmatrix}
        l_1 & l & 1\\
        0 & 0 & 0
    \end{pmatrix}
    \mathcal{E}^{ij}\sum_{k=-2}^{2} T_{ij}^{(k)} A^{(k)}
\nn \\ &
  - \sum_{l_1 = 0}^\infty\sum_{m_l = -l_1}^{l_1}(2l_1 + 1)\i^{-l_1}Y_{l_1}^{*m_l}(0,0)
    \begin{pmatrix}
    l_1 & l & 1 \\
    0 & 0 & 0
    \end{pmatrix}\mathcal{E}^{ij}\sum_{k=-2}^{2} T_{ij}^{(k)}A^{(k)}
    \int_0^R\frac{\d\mu}{\d r}y_{1n}^T(r)j_{l1}(kr)r^2 \d r ,
    \label{eqn:Toroidal_F_sum}
\end{align}
where \(T^{(k)}\) and \(A^{(k)}=A^{(k)}(l_1,m_1,l,m)\) are given in the
appendix. The terms including \(\lambda\) and its derivative involve only
the $A^{(0)}$ term and are given by
\begin{align}
  \frac{\sqrt{l(l+1)}}{4\pi}  F_T^{(\lambda)} & =
 \lambda(R)R^2y_{1n}^T(R)\sum_{l_1=0}^\infty\sum_{m_l = -l_1}^{l_1}(2l_1+1)
    \i^{-l_1}j_{l_1}(k R)Y_{l_1}^{*m_1}(e,\lambda)
    \begin{pmatrix}
        l_1 & l & 1\\
        0 & 0 & 0
    \end{pmatrix}
    A^{(0)}
    \nn\\ & +  
  \sum_{l_1 = 0}^\infty\sum_{m_l = -l_1}^{l_1}(2l_1 + 1)\i^{-l_1}Y_{l_1}^{*m_l}(e,\lambda)
    \begin{pmatrix}
    l_1 & l & 1 \\
    0 & 0 & 0
    \end{pmatrix} A^{(0)}
    \int_0^R\frac{\d\lambda}{\d r}y_{1n}^T(r)j_{l1}(kr)r^2 \d r.
\end{align}
For the case of \(F_T^{(\lambda)}\), the mode of particular interest is the
case \(l=m=2\), as it is the lowest mode capable of oscillations. Taking the
sum over \(l_1\) we observe that \(A^{(0)}\) vanishes for all \(l_1\).
Therefore we need to consider only \(F^{(\mu)}_T\). In $F_T$, we have introduced
Wigner's $6j$ symbols
$\left( \begin{smallmatrix} a&b&c\\ c&d&e \end{smallmatrix} \right)$
which are defined in Eq.~(\ref{Wigner_def}). Assuming a spherical body,
we are free to rotate the coordinate system in such a way to align the
momentum vector of the gravitational wave with the $z$ axis of our coordinate
system, $k_i = \omega\hat{e}^{(z)}_i$.
The choice of coordinate system is also the reason for why
\(Y^{*m_1}_{l_1}(\psi,\chi)\) which depends in general on the angles of the
incoming GW wave  simplifies to \(Y^{*m_1}_{l_1}(0,0)\) in
\eqref{eqn:Toroidal_F_sum}.
The contraction of the polarization tensor with the \(T\) matrix given in the
appendix results in
\begin{equation}
    \mathcal{E}_{ij}\sum_{k=-2}^{2} T^{ij(k)}A^{(k)} = \frac{\i}{2\sqrt{2}}\sqrt{(l+m)(l-m+1)}\begin{pmatrix}
        l_1 & l & 1\\
        m_1 & m-1 & -1
    \end{pmatrix}.\label{eqn:polarisation_contract_simplified}
\end{equation}
Since the Bessel function \(j_l(x)\) satisfies
\begin{equation}
    j_{l}(x) \simeq \frac{1}{(2l+1)!!}(x)^{l}\label{eqn:bessel_approx}
\end{equation}
for $x \to 0$,  we have a series which very quickly converges given that \(x\ll 1\). Inserting for \eqref{eqn:bessel_approx} and \eqref{eqn:polarisation_contract_simplified} in \eqref{eqn:Toroidal_F_sum} gives
\begin{subequations}
    \begin{align}
    \frac{\sqrt{l(l+1)}}{8\pi\i}F_T^{(\mu)} =& \sum_{l_1=0}^\infty\sum_{m_l = -l_1}^{l_1}\i^{-l_1}\frac{(2l_1+1)}{(2l_1+1)!!}\sqrt{(l+m)(l-m+1)}
    \sqrt{\frac{(l_1-m_1)!}{(l_1+m_1)!}}P_{l_1}^{m_1}(1)
    \begin{pmatrix}
        l_1 & l & 1\\
        0 & 0 & 0
    \end{pmatrix}
    \begin{pmatrix}
        l_1 & l & 1\\
        m_1 & m-1 & -1
    \end{pmatrix}\nn\\
    &\times\left[\mu(R)R^2y_{1n}^T(R)(kR)^{l_1}-\int_0^R\frac{\d\mu}{\d r}y_{1n}^T(r)(kr)^{l_1}r^2\d r\right].
\end{align}
\end{subequations}
Together with the time-dependence~\eqref{g_bar} and the
normalization~\eqref{normT}, we now have all the required ingredients to
calculate the toroidal displacement after having determined numerically
the parameter function $y_{1n}(r)$.

\subsubsection{GW induced spheroidal motion}

Following the same strategy as in the toroidal case, we obtain using
\eqref{eqn:spheroidal_displacement_1b} for the induced displacement in the
case of spheroidal oscillations 
\begin{equation}
  u^{(nml)}_j(\vec{r},t) = h_0\bar{g}(t) (\Lambda_S^{nml})^{-1}
  \left[P^{*(nml)}_j(\vec{r}) F_{S_1} + l(l+1)B^{*(nml)}_j(\vec{r}) F_{S_2}\right],
    \label{eqn:chapter4_displacement_spheroidal}
\end{equation}
where \(F_{S_1}\) and  \(F_{S_2}\) are given by
\begin{align}
\sqrt{l(l+1)}  F_{S_1} &= R^2\mu(R)y_{1n}(R) \mathcal{E}^{ij}
    \widetilde P^{(lm)}_{ij}(kR)
  -\int_0^R\frac{\d\mu}{\d r}y_{1n}(r)r^2dr
   \mathcal{E}^{ij}  \widetilde P^{(lm)}_{ij}(kr) ,
\end{align}
\begin{align}
\sqrt{l(l+1)}  F_{S_2} &=R^2\mu(R)y_{1n}(R) \mathcal{E}^{ij}
        \widetilde B^{(lm)}_{ij}(kR)
  -\int_0^R\frac{\d\mu}{\d r}y_{1n}(r)r^2dr
  \mathcal{E}^{ij}  \widetilde B^{(lm)}_{ij}(kr) .
\end{align}\label{eqn:Fs1andFs2}
Note that the terms including the first Lam\'e parameter $\lambda$
do not contribute to spheroidal oscillations, see the appendix for further details. 
In order to simplify the expressions for \(F_{S_1}\) and \(F_{S_2}\),
we will study the (reducible) quadrupole moment ${D}_{ij}$ of the Moon,
\begin{equation}
    \mathcal{D}_{ij} = 3\int_V r_i r_j \rho \d V.
\end{equation}
Here, we kept the trace $\mathcal{D}_{ii}$ which can couple to the
scalar polarization state of the GW.
For a displacement \(r_i\xrightarrow{} r_i + u_i\), the change of quadrupole
moment to first order is
\begin{equation}
    \delta\mathcal{D}_{ij} = 3\int_V (r_i u_j + u_i r_j)\rho \d V.
    \label{eqn:spheroidal_delta_quadrupole}
\end{equation}
If we consider only a single spheroidal mode, then we can represent the displacement contribution to \(\delta\mathcal{D}_{\mu\nu}\) by
\begin{equation}
    u^{(nml)}_{j} = y_{1n}P^{ml}_j+y_{3n}\sqrt{l(l+1)}B^{ml}_j.
    \label{eqn:spheroidal_displacement_from_quadrupole}
\end{equation}
Inserting \eqref{eqn:spheroidal_displacement_from_quadrupole} into \eqref{eqn:spheroidal_delta_quadrupole} we can split the total quadruple moment into 
\begin{subequations}
  \begin{equation}
    \delta\mathcal{D}^{(P)}_{ij} = 3\int_0^Ry_{1n}(r)\rho^{(0)}(r)r^2\d r
    \left(\widetilde P^{(ml)}_{ij}(r)
     + \widetilde P^{(ml)}_{ji}(r)  \right) ,\\
    \end{equation}
    \begin{equation}
      \delta\mathcal{D}^{(B)}_{ij} = 3\sqrt{l(l+1)}
      \int_0^Ry_{3n}(r)\rho^{(0)}(r)r^2\d r
      \left( \widetilde B^{(ml)}_{ij}(r) + \widetilde B^{(ml)}_{ji}(r)\right) .
    \end{equation}\label{eqn:chapter4_delta_quadrupole}
\end{subequations}
Neglecting the phase factor, we can express $F_{S_1}$ and $F_{S_2}$ via the
changes of the quadruple moment as 
\begin{align}
    F_{S_1} &= \frac{R^2\mu(R)y_{1n}(R)-\int_0^R\Dot{\mu}y_{1n}(r)r^2\d r}{3\int_0^R\rho_0y_{1n}(r)r^2\d r} \,\mathcal{E}^{ij}\delta\mathcal{D}^{(P)}_{ij},\\
    F_{S_2} &= \frac{R^2\mu(R)y_{3n}(R)-\int_0^R\Dot{\mu}y_{3n}(r)r^2\d r}{3\int_0^R\rho_0y_{3n}(r)r^2\d r}\,\mathcal{E}^{ij}\delta\mathcal{D}^{(B)}_{ij}.
\end{align}
Thus these functions are proportional to the  interaction
between the GW and the reducible quadrupole moment of the
spherical body.

We employ as in the toroidal case the low-frequency approximation.
Starting from Eqs.~(\ref{fourierP}) and (\ref{fourierB}) for
the Fourier transformed surface harmonics derived in the appendix,
we consider the limit $kr\to 0$. Then the Bessel function simplifies to
\(j_{l_1}(0) = \delta_{l_1,0}\). Forcing \(l_1=0\) for a nonzero result puts
strict restrictions on the Wigner symbols in \(D^{(j)}\) and the integrals
simplify to
 \begin{subequations}
   \begin{equation}
    \widetilde P^{(lm)}_{ij}(0)=
   \frac{4\pi}{5}\delta_{l,2}(\Gamma_{ij}^{(0)}\delta_{m,0} -\Gamma_{ij}^{(1)}\delta_{m,-1}+\Gamma_{ij}^{(-1)}\delta_{m,1}+\Gamma_{ij}^{(2)}\delta_{m,-2}+\Gamma_{ij}^{(-2)}\delta_{m,2}),
\end{equation}
   \begin{equation}
     \sqrt{l(l+1)} \widetilde B^{(lm)}_{ij}(0)=
     \frac{12\pi}{5}\delta_{l,2}(\Gamma_{ij}^{(0)}\delta_{m,0} -\Gamma_{ij}^{(1)}\delta_{m,-1}+\Gamma_{ij}^{(-1)}\delta_{m,1}+\Gamma_{ij}^{(2)}\delta_{m,-2}+\Gamma_{ij}^{(-2)}\delta_{m,2}).
   \end{equation}
   \label{kr0}
 \end{subequations}
If we insert this into the equations~\eqref{eqn:chapter4_delta_quadrupole} for
the change in the quadrupole moment, we obtain
\begin{equation}
  \delta\mathcal{D}_P + \delta\mathcal{D}_B =
  \frac{24\pi}{5\sqrt{6}}\delta_{l,2}\Delta\int_0^R(y_{1n}+3y_{3n})\rho_0(r)r^2\d r
  \label{eqn:quadrupole_sum}
\end{equation}
with
\begin{equation*}
    \Delta = \begin{pmatrix}
        \delta_{m,2} + \delta_{m,-2} - \sqrt{\frac{2}{3}}\delta_{m,0} & \i\delta_{m,2}-\i\delta_{m,-2} & \delta_{m,1}+\delta_{m,-1}\\
        \i\delta_{m,2}-\i\delta_{m,-2} & -\delta_{m,2} - \delta_{m,-2} - \sqrt{\frac{2}{3}}\delta_{m,0} &
        -\i\delta_{m,-1}+\i\delta_{m,1}\\
        \delta_{m,1} + \delta_{m,-1} &
        -\i\delta_{m,-1}+\i\delta_{m,1} &
        2\sqrt{\frac{2}{3}}\delta_{m,0}
    \end{pmatrix}.
\end{equation*}

The quadrupole tensor has to be contracted with the polarization tensor of the
GW.  Because of the Kronecker delta \(\delta_{l,2}\), only spheroidal
oscillations with \(l=2\) will contribute in this approximation. The contribution of
the scalar polarization state is given by
\begin{equation}
    h_s(\delta_{m,2}+\delta_{m,-2}-\sqrt{\frac{2}{3}}\delta_{m,0}) +h_s(-\delta_{m,2}-\delta_{m,-2}-\sqrt{\frac{2}{3}}\delta_{m,0}) = -2h_s\sqrt{\frac{2}{3}} ,
\label{ratio1}
  \end{equation}
while the standard TT polarizations from Einstein gravity
results in
\begin{equation}
  2h_+(\delta_{m,2} + \delta_{m,-2}),\quad\quad\text{ and } \quad\quad
  2\i h_\times(\delta_{m,2} - \delta_{m,-2}).
\label{ratio2}
\end{equation}
\end{widetext}

\subsection{Numerical  determination of $y_i(r)$}

The last ingredient needed to calculate the  displacement of the Moon surface
are the parameter functions $y_{1n}(r)$. For their  numerical determination,
we follow the  procedure described in chapter~6.7 of
Ref.~\cite{Ben-Menahem:1983}.
In the toroidal case, we impose the boundary
conditions~\eqref{eqn:toroidal_numerical_bc} on the system of differential
equations~\eqref{eqn:toroidal_numericall_diff_eqs}. A guess for \(y_1\) at the
core-mantle boundary $r=a$ and for the eigenfrequency \(\omega\) is used to
begin the integration. The eigenfrequency is then adjusted after the
integration until the second boundary condition for \(y_2\) is satisfied.
Lastly \(y_1\) is normalized. 
In the spheroidal case, the numerical procedure for solving the system of
differential equation system is more complicated, since we have to
consider also the liquid core. The integration
is therefore split into two systems: One differential equation system for
the core, \eqref{eqn:spheroidal_numerical_diff_eqns_core}, and one system for
the mantle, \eqref{eqn:spheroidal_numerical_diff_eqns_mantle}. More
specifically, two initial conditions are chosen for \(y_2\) and \(y_6\)
at \(r=0\), while the remaining initial conditions are set to zero.
We use a Runge-Kutta solver to integrate the
system~\eqref{eqn:spheroidal_numerical_diff_eqns_core} over the core.
The end values are used as initial values for the second integration over
the mantle, except for \(y_3\), where the initial condition is chosen
similarly to \(y_2\) and \(y_6\) at the start. Integration is then done
from \(r=a\) to \(r=R\). We perform this integration three times with
differently chosen initial conditions for \(y_2\), \(y_6\) and \(y_3\).
Then we construct the matrix
$$
    \begin{bmatrix}
    y_2^{(1)} & y_2^{(2)} & y_2^{(3)}\\
    y_4^{(1)} & y_4^{(2)} & y_4^{(3)}\\
    y_6^{(1)} + \frac{l(l+1)}{R}y_5^{(1)} & y_6^{(2)} + \frac{l(l+1)}{R}y_5^{(2)} & y_6^{(3)} + \frac{l(l+1)}{R}y_5^{(3)}
\end{bmatrix}_{r=R} .
$$
The eigenfrequency is varied until the determinant of the matrix changes
sign. The zero determinant of the matrix signals that all boundary conditions
are satisfied and the eigenfrequency is found.

In order to test our numerical procedure, we applied it to the
Jeffreys-Bullen A' model and found good agreement with the eigenfrequencies
and $y_i$ functions described in Ref.~\cite{SeismicWaveAndSources}, for
more details on the numerical implementation and the tests see
Ref.~\cite{Noedtvedt23}.

\section{Lunar response to gravitational waves}

\subsection{Moon models}

\begin{figure}[htp]
\centering
\subfigure{\includegraphics[width=0.85\columnwidth]{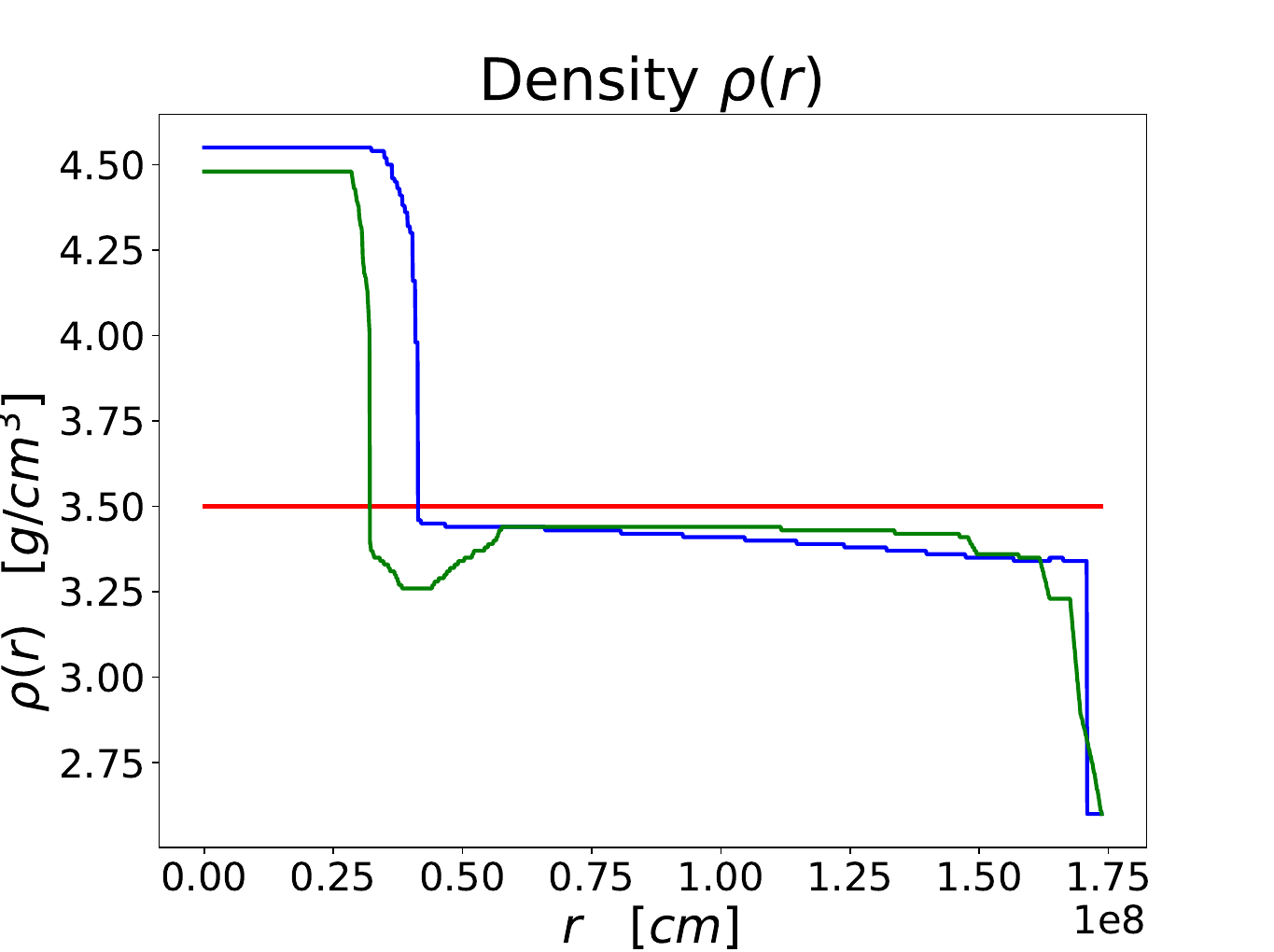}}
\subfigure{\includegraphics[width=0.85\columnwidth]{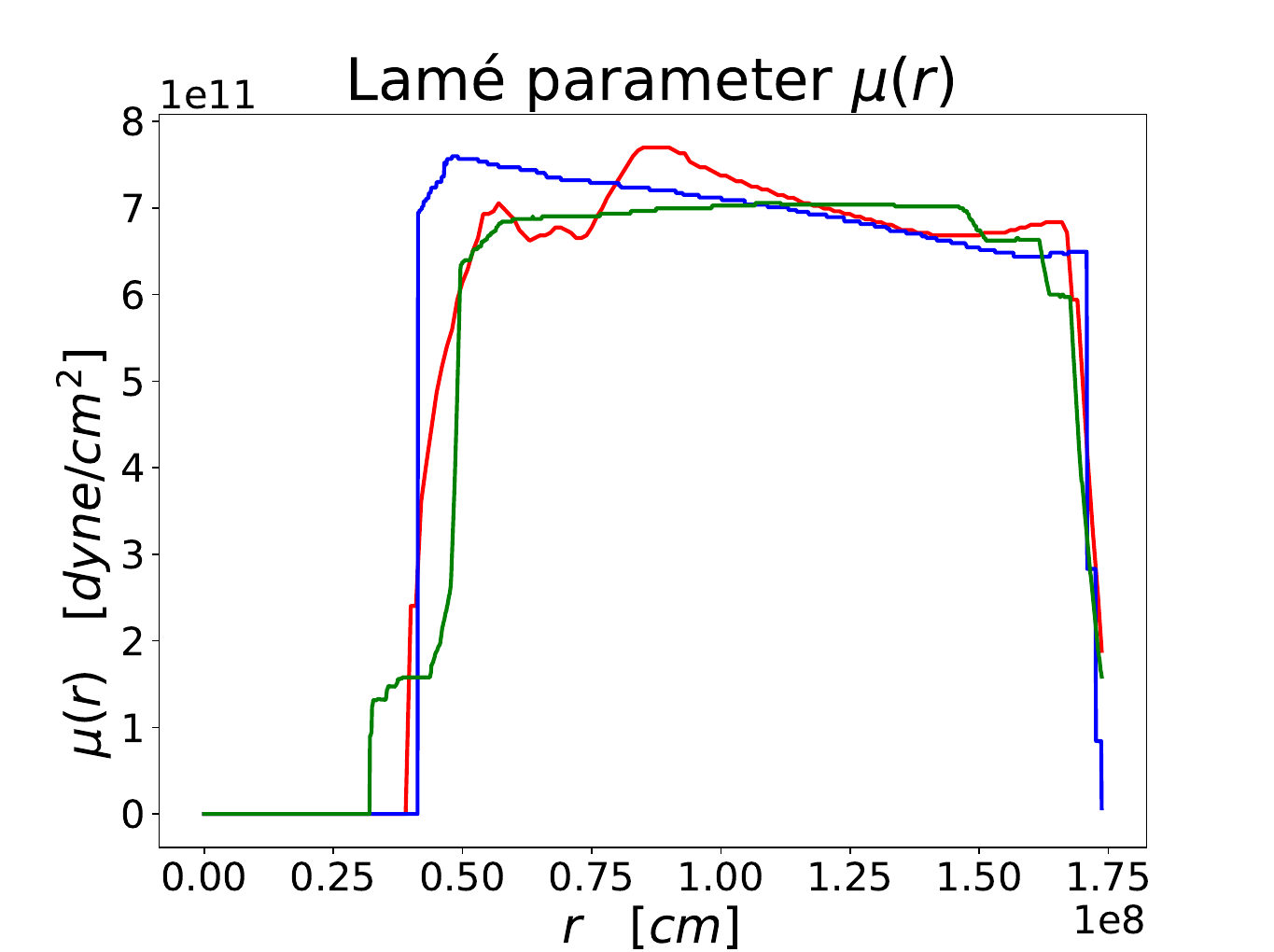}}\\
\subfigure{\includegraphics[width=0.85\columnwidth]{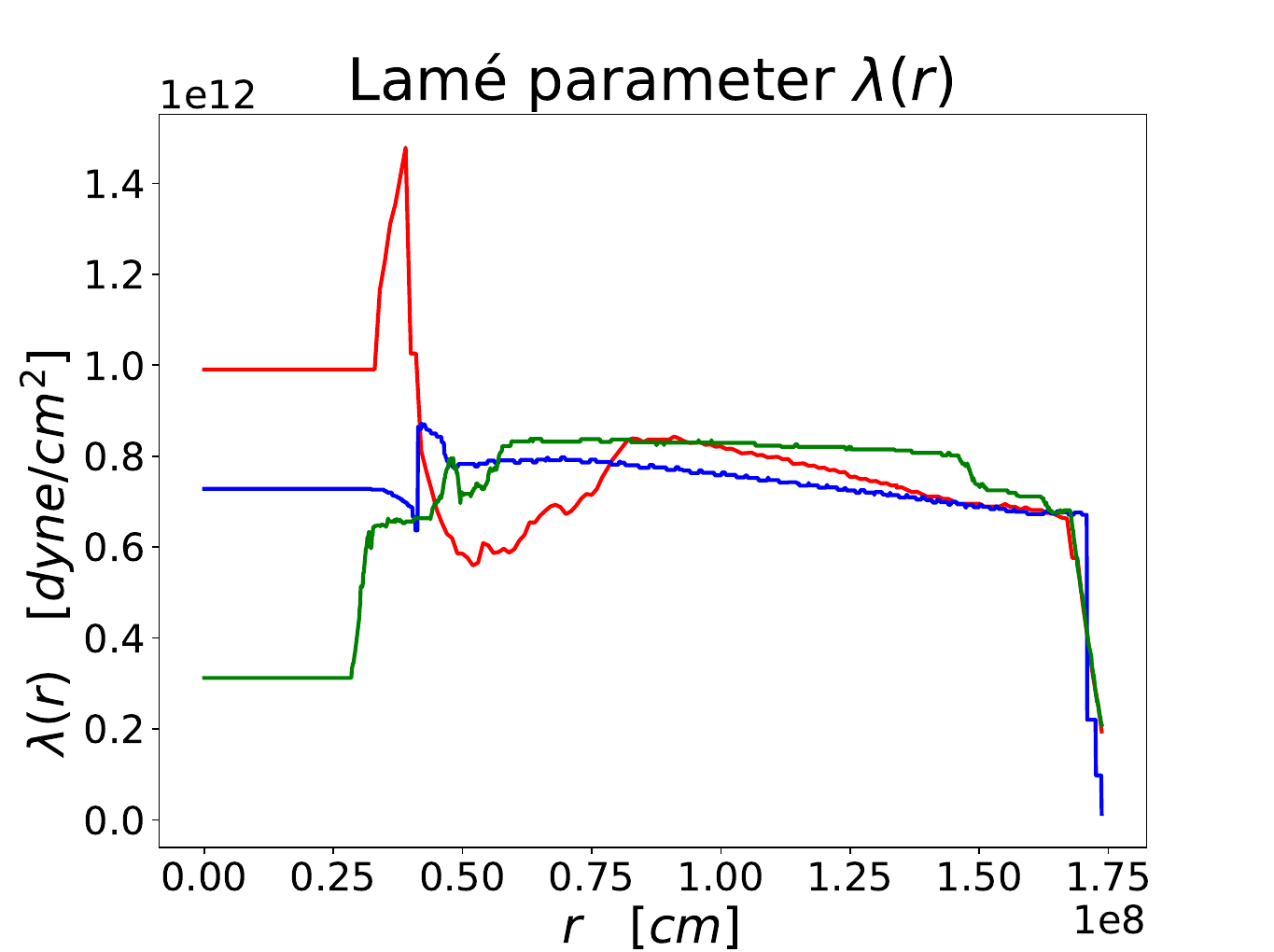}}
\subfigure{\includegraphics[width=0.85\columnwidth]{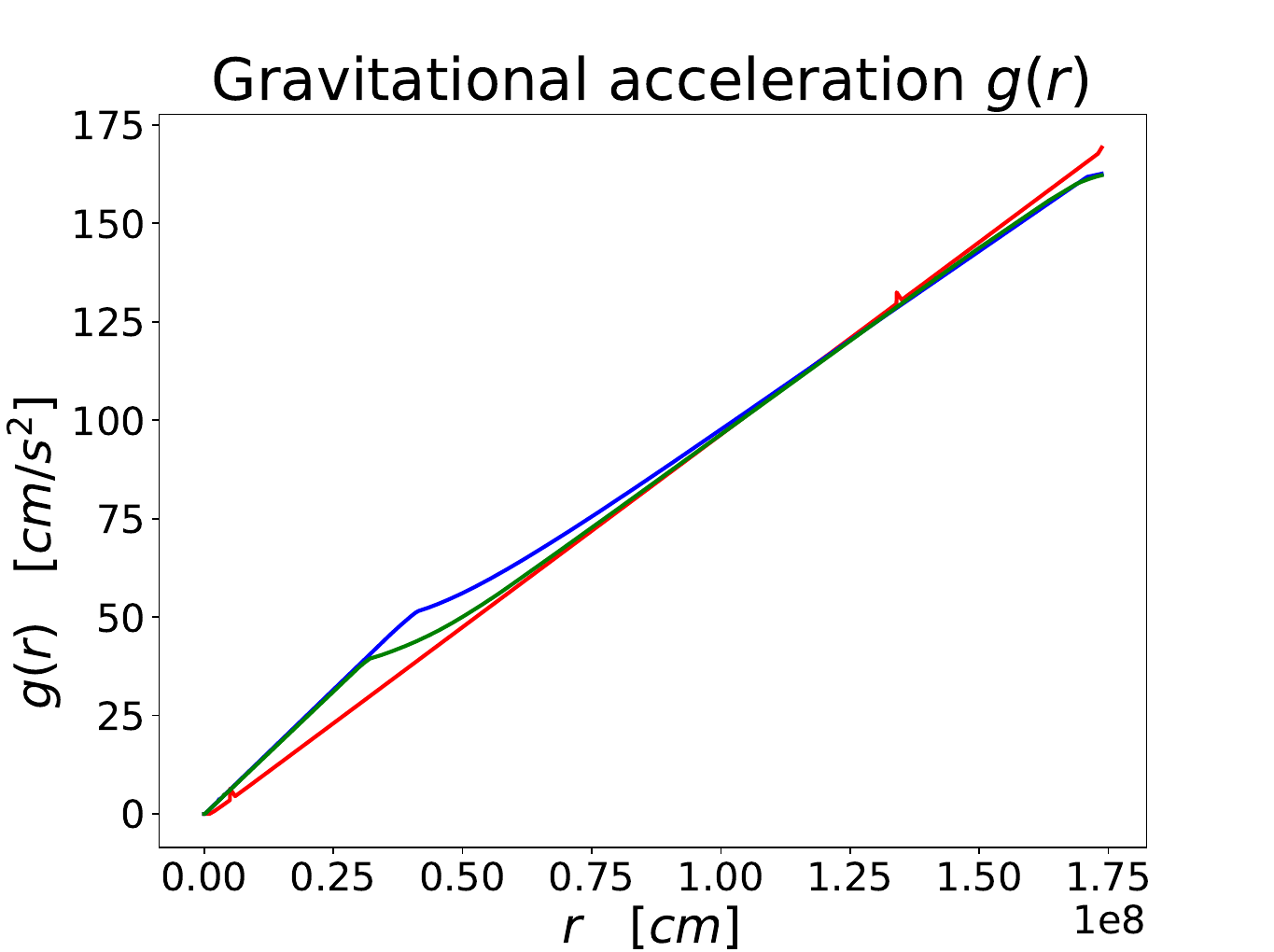}}\\
\caption{The parameters \(\rho(r)\), \(\mu(r)\), \(\lambda(r)\) and \(g(r)\) of the three Moon models. Model 1 in red, model 2 in blue and model 3 in green.}
\label{fig:MoonModelsParameterPlots}
\end{figure}

We use the three different models for the interior of the Moon presented 
in  Ref.~\cite{GarciaLS:2019} to analyze its response to GWs. The key
characteristics of each model, the density \(\rho(r)\), the Lamé parameters
\(\lambda(r)\) and \(\mu(r)\), and the gravitational acceleration \(g(r)\),
are shown in Fig.~\ref{fig:MoonModelsParameterPlots}.
The three models agree well with each-other in most of the mantle, while in
the core deviations are stronger. Moreover, the models slightly
disagree on the value of the core radius $a$. The deviations are largest for
the first Lamé parameter \(\lambda(r)\) in the core.

\subsection{Eigenfrequencies and displacement}

\subsubsection{Eigenfrequencies and normal modes}

The first four eigenfrequencies $\nu_i=\omega_i/2\pi$ of the three models
are shown
in Table~\ref{table:MoonModelsEigenfrequencys}, for toroidal oscillations
on the top and spheroidal oscillations on the bottom.
The fundamental eigenfrequencies of all three models agree very well,
with deviations in the promille range, despite rather large model
differences especially in the core.
The variation of the eigenfrequencies between the three models increases
with $n$, reaching already 20\% for $n=4$. Similarly, the differences
in the eigenfrequencies between toroidal and spheroidal oscillations are small
for $n=1$, and increase with $n$.

\begin{table}[tb]
\centering
    \begin{tabular}{c|c|c|c} 
    $n$ & Model 1 & Model 2 & Model 3 \\ \hline
      1    &  0.993   &  1.014 &  1.011  \\
      2    &  2.901  &  2.950 &  2.918    \\
      3    &  4.266  &  4.388 &  4.079    \\
      4    &  5.635  &  5.869 &  4.929    \\ \hline
      1    &  1.020  & 1.050  &  1.047   \\
      2    &  1.848  & 1.890  &  1.892    \\
      3    &  2.932  & 2.881  &  2.686    \\
      4    &  3.976  &  3.993 &  3.412   
    \end{tabular}
    \caption{The first four  toroidal (top) and spheroidal (bottom)
      eigenfrequencies $\nu_i$ in ms$^{-1}$  of the three Moon models}
\label{table:MoonModelsEigenfrequencys}
\end{table}


In the case of spheroidal oscillations ${}_m S_l$, the leading contribution
in the long-wavelength limit $kr\to 0$ is given by $l_1=0$. Taking into
account the condition~(\ref{kr0}) in Eq.~(\ref{eqn:quadrupole_sum}), we have
seen that then only $l=\pm 2$ modes contribute to  spheroidal oscillations.
Thus we have to consider only the ${}_2 S_2$ mode for the cross and
plus polarization, and  the ${}_0 S_2$ mode for the scalar polarization
state.
In the case of toroidal oscillations, the $l_1=0$ term vanishes and the
leading contribution is thus given by $l_1=1$. In this case,
we have to consider only  the ${}_2 T_2$ mode for
the possible three polarization states of the GW.

\subsubsection{Displacement}

Since the perturbations of a spherically symmetric body factorize in an
angular and a radial dependent part, it is useful to split the displacement
into a part characterizing the overall magnitude of the mode,
\begin{align}
  \xi_T(t) & = h_0(\Lambda_T^{022})^{-1} F_T\bar{g}(t),\label{eqn:xi_toroidal}
             \\
    \xi_S(t)& = h_0(\Lambda_S^{022})^{-1}(F_{S_1}+F_{S_2})\bar{g}(t),\label{eqn:xi_spheroidal}
\end{align}
neglecting the angular-dependent modulation determined by the
$\widetilde V$ functions.
For the time-dependence  $\bar g_n(t)$ we use Eq.~(\ref{barg}) where 
we assume as quality factor \(Q_1 = 3300\) for the first eigenfrequency
for all Moon models, following Ref.~\cite{GarciaLS:2019}. Moreover, we consider
first the displacement of the models at resonance, i.e. when \(\omega_0 =
\omega_1\), where \(\omega_0\) is the frequency of the GW.

\begin{figure}[htp]
\centering
\subfigure{\includegraphics[width=\columnwidth]{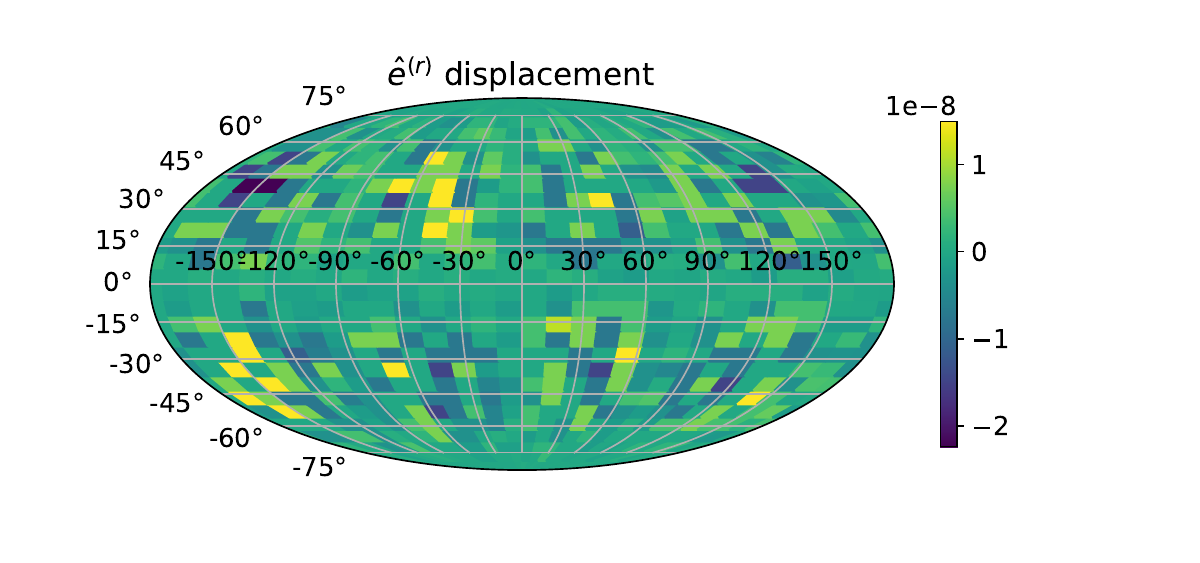}}
\subfigure{\includegraphics[width=\columnwidth]{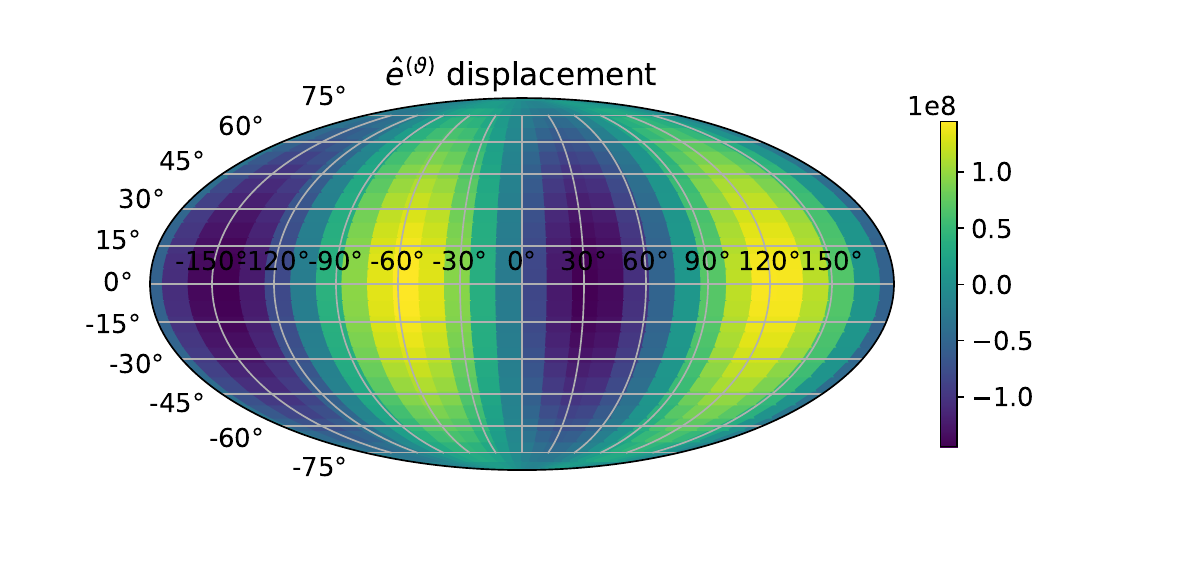}}
\subfigure{\includegraphics[width=\columnwidth]{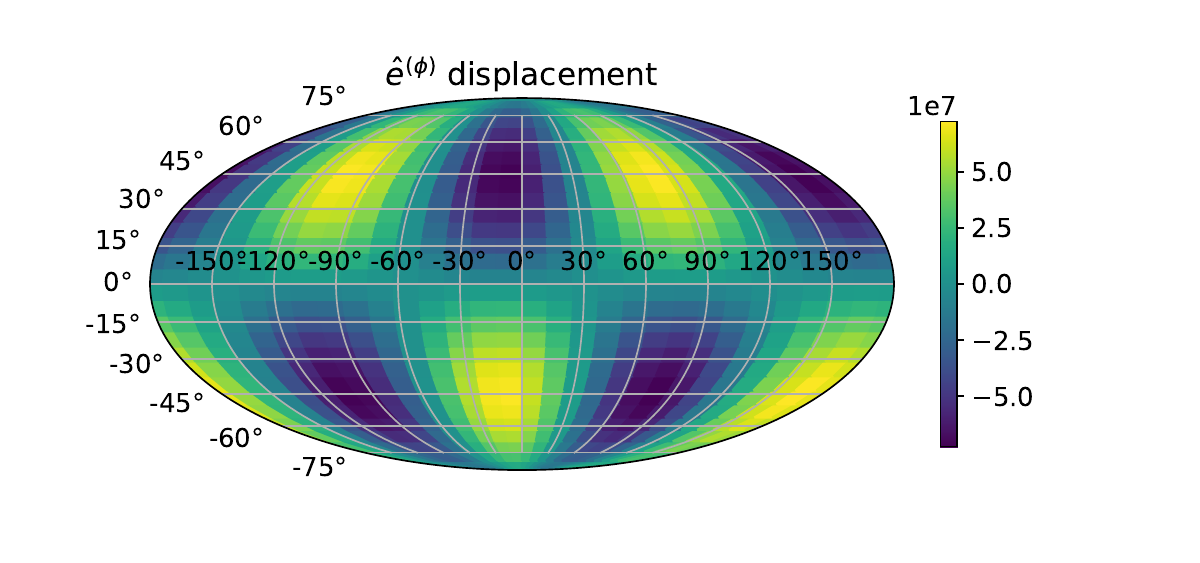}}
\caption{The \(\tensor[_2]{T}{_2}\) displacement per unit strain \(h_0 = 1\) of Moon model 1 to a plus and cross polarized  GW in units of cm. The top figure is the displacement in the \(\hat{e}^{(r)}\) direction, the middle figure in the \(\hat{e}^{(\theta)}\) direction and the bottom figure in the \(\hat{e}^{(\phi)}\) direction.}
\label{fig:ToroidalDisplacement_M1_T22}
\end{figure}

We begin with the toroidal response. In
Fig.~\ref{fig:ToroidalDisplacement_M1_T22}, we show
the toroidal displacement $\xi_T$ for the \(\tensor[_2]{T}{_2}\) mode using
the Moon model~1 and setting $h_0=1$. Moreover, we have here assumed
\(A_+ =A_\times = 1\). The shape of the oscillation pattern is
independent
of the specific model, only the magnitude \(\xi_T\) varies.  The values
of \(\xi_T\) determined by our numerical integrations are summarized in
Table~\ref{table:MoonModelsCompareResponse}. 
The toroidal displacement of the Moon is typically  two orders of magnitude
larger than for the Earth. There is only a minor difference between
model~1 and 3, while the response in model~2 is a factor five smaller.
Looking back at Fig.~\ref{fig:MoonModelsParameterPlots}, we do observe that
model~2 differs from the other two models close to the surface, particularly
in the second Lam\'e parameter \(\mu(r)\): This parameter is in model~2
two orders lower than in Model~1 and 3.

\begin{figure}[htp]
\centering
\subfigure{\includegraphics[width=\columnwidth]{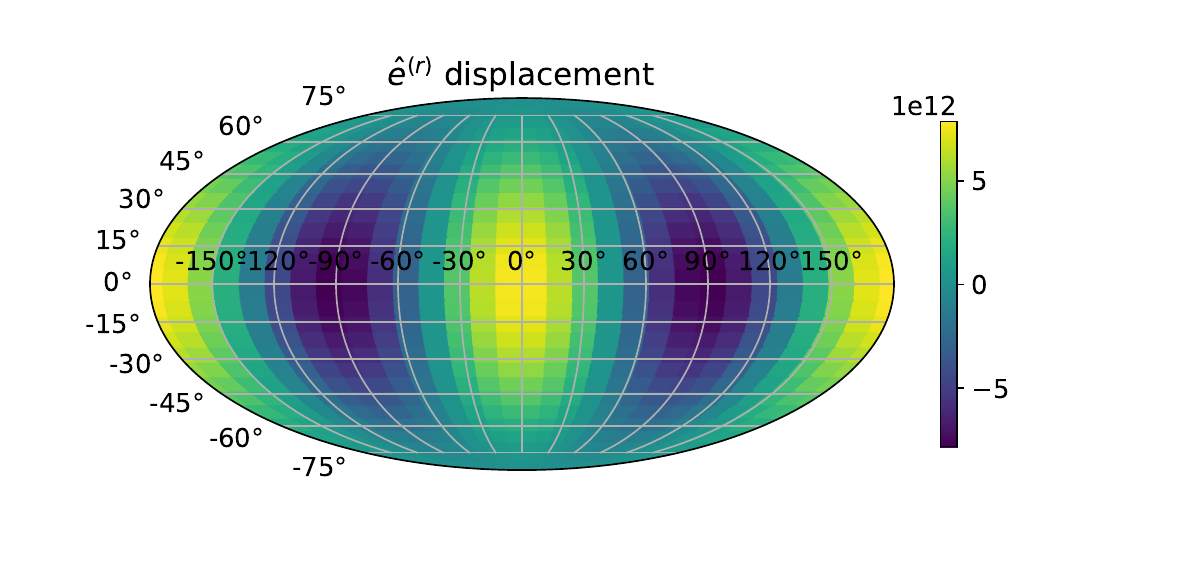}}
\subfigure{\includegraphics[width=\columnwidth]{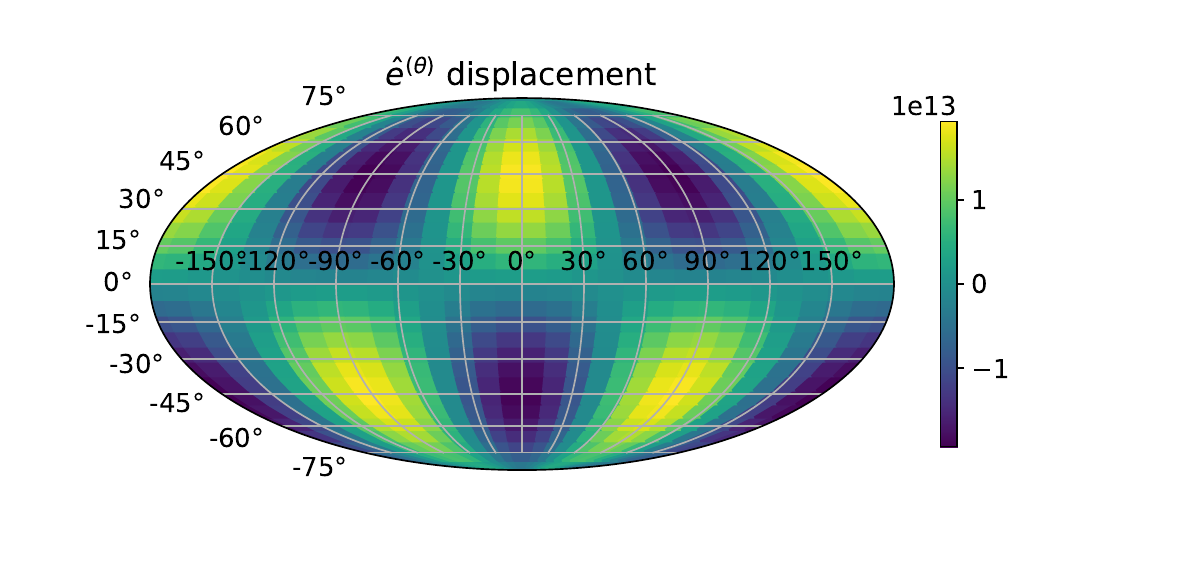}}
\subfigure{\includegraphics[width=\columnwidth]{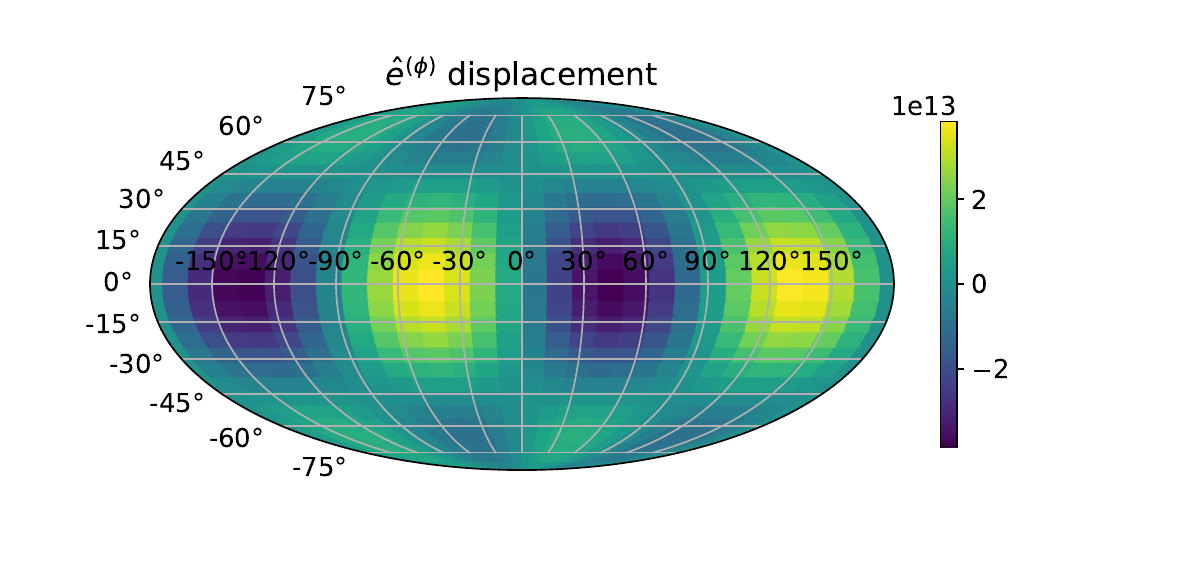}}
\caption{The \(\tensor[_2]{S}{_2}\) displacement of Moon model 1 to a plus- and cross-polarized gravitational wave in units of cm. The top figure shows the displacement in the \(\hat{e}^{(r)}\) direction, the middle figure shows the displacement in the \(\hat{e}^{(\theta)}\) direction and the bottom figure shows the displacement in the \(\hat{e}^{(\phi)}\) direction.} 
\label{fig:SpheroidalDisplacement_M1_S22}
\end{figure}

We proceed to the displacement from the spheroidal oscillations.
In Fig.~\ref{fig:SpheroidalDisplacement_M1_S22}, we show the displacement
for the \(\tensor[_2]{S}{_2}\) mode in Model~1. Comparing the values for
$\xi_S$ given for the three Moon models in Table~\ref{table:MoonModelsCompareResponse}, we do observe the opposite trend as for the toroidal mode in that the
response of Model~2 is larger than in the other models. Also, 
the three models agree better than in the toroidal case.
This different behavior may arise from the spheroidal response depending also
on the parameter values of the core and not just the mantle as in the
toroidal case.

\begin{figure}[htp]
\centering
\subfigure{\includegraphics[width=\columnwidth]{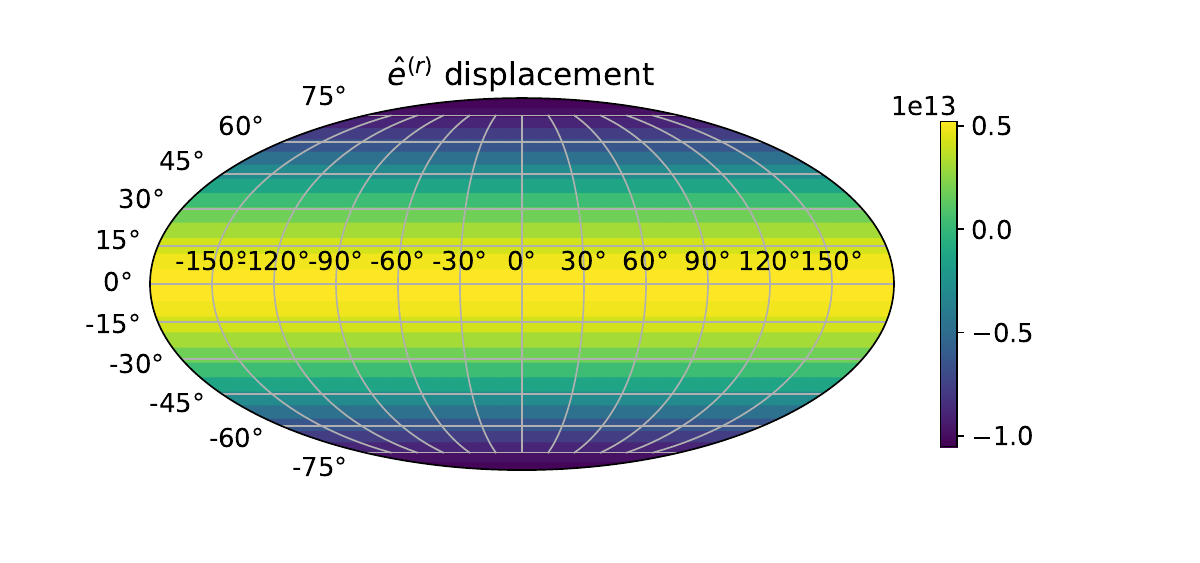}}
\subfigure{\includegraphics[width=\columnwidth]{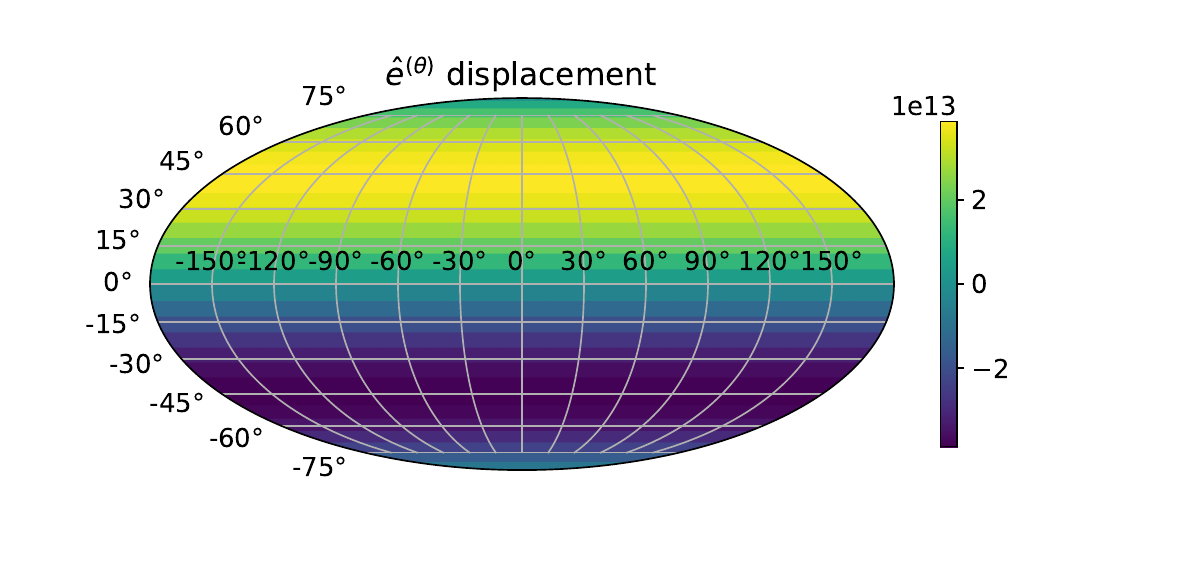}}
\subfigure{\includegraphics[width=\columnwidth]{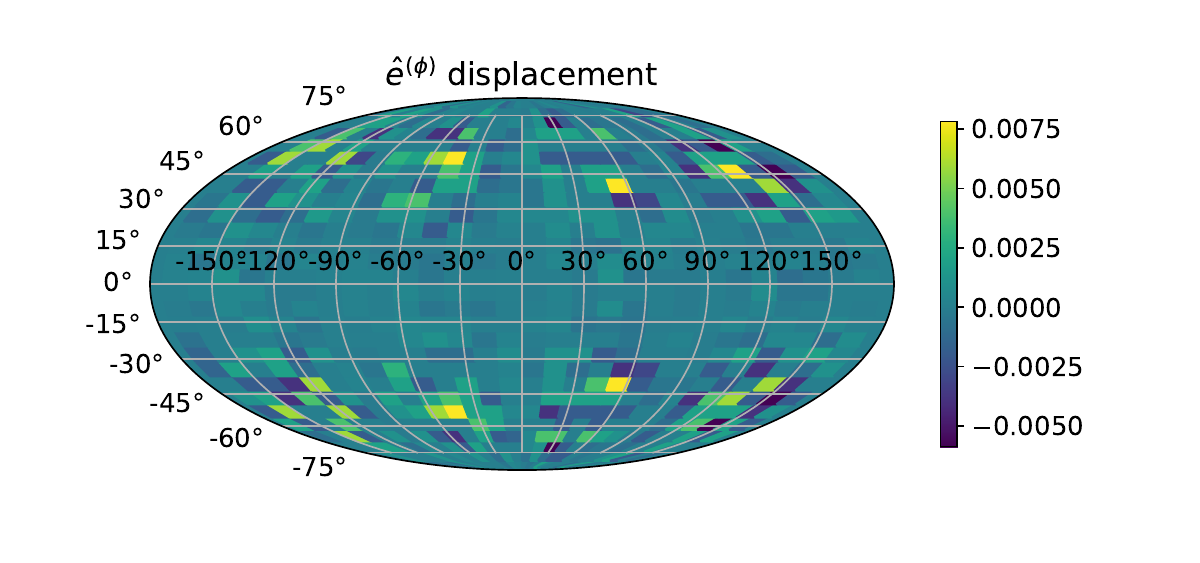}}
\caption{The \(\tensor[_0]{S}{_2}\) displacement of Moon model 1 to a scalar polarized gravitational wave in cm. The top model shows the displacement in the \(\hat{e}^{(r)}\) direction, the middle model shows the displacement in the \(\hat{e}^{(\theta)}\) direction and the bottom model shows the displacement in the \(\hat{e}^{(\phi)}\) direction.}
\label{fig:SpheroidalDisplacement_BD_M1_S22}
\end{figure}

The last mode for discussion is the mode excited by the scalar polarized
gravitational wave \(\tensor[_0]{S}{_2}\). The response for the first model is
plotted in spherical coordinates in Fig.~\ref{fig:SpheroidalDisplacement_BD_M1_S22}. The values of \(\xi_S\) follow a similar
pattern, with the largest response in model 2, while the weakest response
happens in model 1. For all models, the response to the scalar mode  \(\tensor[_0]{S}{_2}\)is smaller than for  the \(\tensor[_2]{S}{_2}\) modes. The ratio
of the displacement for \(\tensor[_0]{S}{_2}\) and \(\tensor[_2]{S}{_2}\)
modes is in agreement with the factor $\sqrt{2/3}$ difference found in 
Eqs.~(\ref{ratio1}) and ~(\ref{ratio2}).

In addition to the resonance case, we report in
  Table~\ref{table:MoonModelsCompareResponseLower}  the response
at a frequency inbetween the two first eigenfrequencies, which we choose as
\(\omega_0 = (\omega_1+\omega_2)/2\). At such an intermediate frequency,
where the response should be less dependent on the specific choice of the
Green function $G_n(\omega)$, the response is reduced, what can also be
seen clearly from Fig.~\ref{fig:GWResponsePerUnitStrain_OldGreens}.

We can make a crude estimate of the magnitude of the scalar amplitude
expected for the GW signal from a Galactic neutron star.
Following Ref.~\cite{Verma:2021nbz}, the amplitude of a plus-polarised
gravitational wave from a slightly perturbed rotating neutron star can be
expressed as
\begin{equation}
    h_+(t) = \frac{16\pi^2G}{c^4}(1-\zeta)Q\frac{f_0^2}{r}\cos{2\phi(t)},
\end{equation}
where \(\zeta\) is the Brans-Dicke parameter, $Q$ the quadrupol moment, and
\(f_0\) and $\phi$ the rotational frequency and angle of the star.
Still following \cite{Verma:2021nbz}, the scalar polarisation can be
expressed as
\begin{equation}
    h_s(t) = -\frac{4\pi G}{c^3}\zeta\left(D\frac{f_0}{r}\sin{\phi(t)}-\frac{4\pi}{c}Q\frac{f_0^2}{r}\cos{2\phi(t)}\right)
\end{equation}
with $D$ as the stellar dipole moment. 
Choosing \(\phi(t)=\pi/2\), we obtain as estimate for the ratio of the
scalar and plus-polarised response
\begin{equation}
    \frac{h_s}{h_+} = -\frac{\zeta}{1-\zeta}\left(\frac{c}{4\pi f_0}\frac{D}{Q}+1\right).  
\end{equation}
Reference~\cite{Will_2014} reported the bound \(\zeta < 1.25\times 10^{-5}\).
We can make a crude estimate of the ratio $h_s/h_+$ by considering
\(f_0 = 100\)\,Hz and \(Q=10^{33}\)\,\SI{}{\kilogram\meter\squared}.
Moreover, we choose  $D=10^{29}$\,\SI{}{\kilogram\meter} and $D=0$
to bracket the range for the stellar dipole moment. With these values,
we arrive at the following estimates
\begin{equation}
    \frac{h_s}{h_+} \lesssim 3\times 10^{-4} \quad\text{ or }\quad \frac{h_s}{h_+} \lesssim 1\times 10^{-5}
\end{equation}
for a nonzero and zero dipole moment, respectivly.

\begin{table}[htp]
\centering
    \begin{tabular}{c|c|c|c} 
    \toprule
    \toprule
    Mode & Model 1 & Model 2 & Model 3 \\
    \midrule\hline
    \(\tensor[_2]{T}{_2}\)    & $4.478\times10^{7}$\,cm   & $9.612\times10^{7}$\,cm  & $4.722\times10^{7}$\,cm                   \\[1.3ex]
    \(\tensor[_2]{S}{_2}\)    & $1.288\times10^{13}$\,cm  & $1.900 \times10^{13}$\,cm  & $3.636 \times10^{12}$ \,cm                  \\[1.3ex]
    \(\tensor[_0]{S}{_2}\)    & $1.051\times10^{13}$\,cm   & $1.549\times10^{13}$\,cm  & $2.969\times10^{12}$ \,cm                     \\[1.3ex]
    \midrule
    \bottomrule
    \end{tabular}
\caption[]{The displacement \(\xi_T/h_0\) and \(\xi_S/h_0\) for the different Moon models and the \(\tensor[_2]{T}{_2}\), \(\tensor[_2]{S}{_2}\) and \(\tensor[_0]{S}{_2}\) mode for $\omega_0=\omega_1$.}
\label{table:MoonModelsCompareResponse}
\end{table}
\begin{table}[htp]
\centering
    \begin{tabular}{c|c|c|c} 
    \toprule
    \toprule
    Mode & Model 1 & Model 2 & Model 3 \\
    \midrule\hline
    \(\tensor[_2]{T}{_2}\)    & $0.803\times 10^{4}$\,cm   & $1.774\times10^{4}$\,cm  & $0.857\times10^{4}$\,cm                   \\[1.3ex]
    \(\tensor[_2]{S}{_2}\)    & $3.529\times10^{10}$\,cm  & $ 4.574\times10^{10}$\,cm  & $3.369\times10^{10}$\,cm                  \\[1.3ex]
    \(\tensor[_0]{S}{_2}\)    & $2.881\times10^{10}$\,cm   & $ 3.735\times10^{10}$\,cm  & $2.751\times10^{10}$\,cm                     \\[1.3ex]
    \midrule
    \bottomrule
    \end{tabular}
\caption[]{The displacement \(\xi_T/h_0\) and \(\xi_S/h_0\) for the different Moon models and the \(\tensor[_2]{T}{_2}\), \(\tensor[_2]{S}{_2}\) and \(\tensor[_0]{S}{_2}\) mode for \(\omega_0 = (\omega_1+\omega_2)/2\).}
\label{table:MoonModelsCompareResponseLower}
\end{table}

\subsubsection{Total response over frequency}

We have up to this point kept our focus mainly on the first eigenfrequency of
the Moon models, having presented the response of the Moon only for this
frequency.  Since the frequency of the GW will in general not
match the eigenfrequency of the Moon, or may cover a broad range of
frequencies, it is necessary to study how the response changes as we move
to other frequencies for the incoming GW.
If we again assume a GW with the momentum vector
\(\vec{k} = (0, 0, \omega_0)\), then the expression for the main contribution
from spheroidal oscillations is
\begin{equation}
    u_i^{(nml)}(\vec{r},t) = h_0(\Lambda^{(nml)})^{-1}\Bar{g}(t){Q}^{*(nml)}_i(\vec{r})(F_{S_1} + F_{S_2}).
\end{equation}
We set now \(h_0=1\), considering the response per unit strain, and neglect
the angular dependence as well, defining
\(\xi^{nml}(t) = (\Lambda^{(nml)})^{-1}\Bar{g}(t)(F_{S_1} + F_{S_2})\).
We are interested in the total response and we must therefore sum over
eigenfrequencies. We restrict the analysis to the \(l=2\) and \(m=2\)
modes,
\begin{equation} 
    \xi_{\rm tot}(t) = \sum_{n=1}^\infty\xi^{n22}(t).
\end{equation}
We choose the time \(t\) such that the response is maximal and assume
that the response from all the eigenfrequencies adds constructively.
Moreover, we consider as a signal a finite monochromatic wave with
duration \(\tau=T_1\), where \(T_1=2\pi/\omega_1\) is the eigenperiod of
the first eigenfrequency.

\begin{figure}[htp]
\centering
\includegraphics[width=0.49\textwidth]{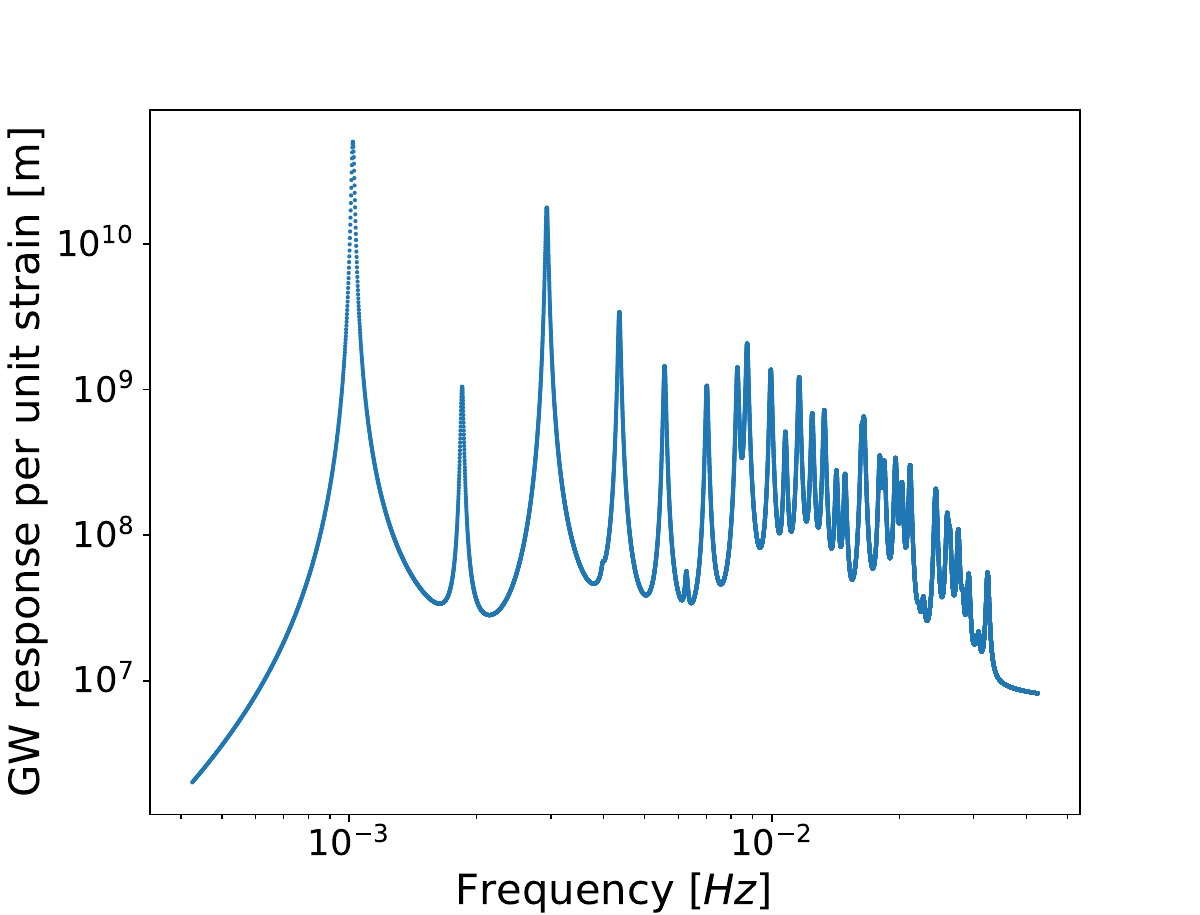}
\caption{The gravitational wave response in Moon model 1 including the first
37 eigenfrequencies with  $Q_n={\rm const.}$ using the response function given in Eq.~(\ref{barg}).}
\label{fig:GWResponsePerUnitStrain_OldGreens}
\end{figure}

We show the gravitational response per unit strain as a function of the
frequency of the incoming GW in
Fig.~\ref{fig:GWResponsePerUnitStrain_OldGreens}.
We choose again as quality factor of \(Q_0 = 3300\) for the first
eigenfrequency. The frequency dependence of the quality factors $Q_n$
for higher eigenmodes of the Moon is rather uncertain:  The authors
of Ref.~\cite{Nakamura92} found  $Q_n \propto\omega^{0.7}$ in the
range 3--8\,Hz for $S$ waves, while for $P$ waves no significant
frequency dependence was found. As there are no determinations of
the frequency dependence of $Q$ in the most interesting range mHz--Hz
range, we choose $Q_n$ a constant, \(Q_n = Q_0=3300\).

\begin{figure}[htp]
\centering
\includegraphics[width=0.49\textwidth]{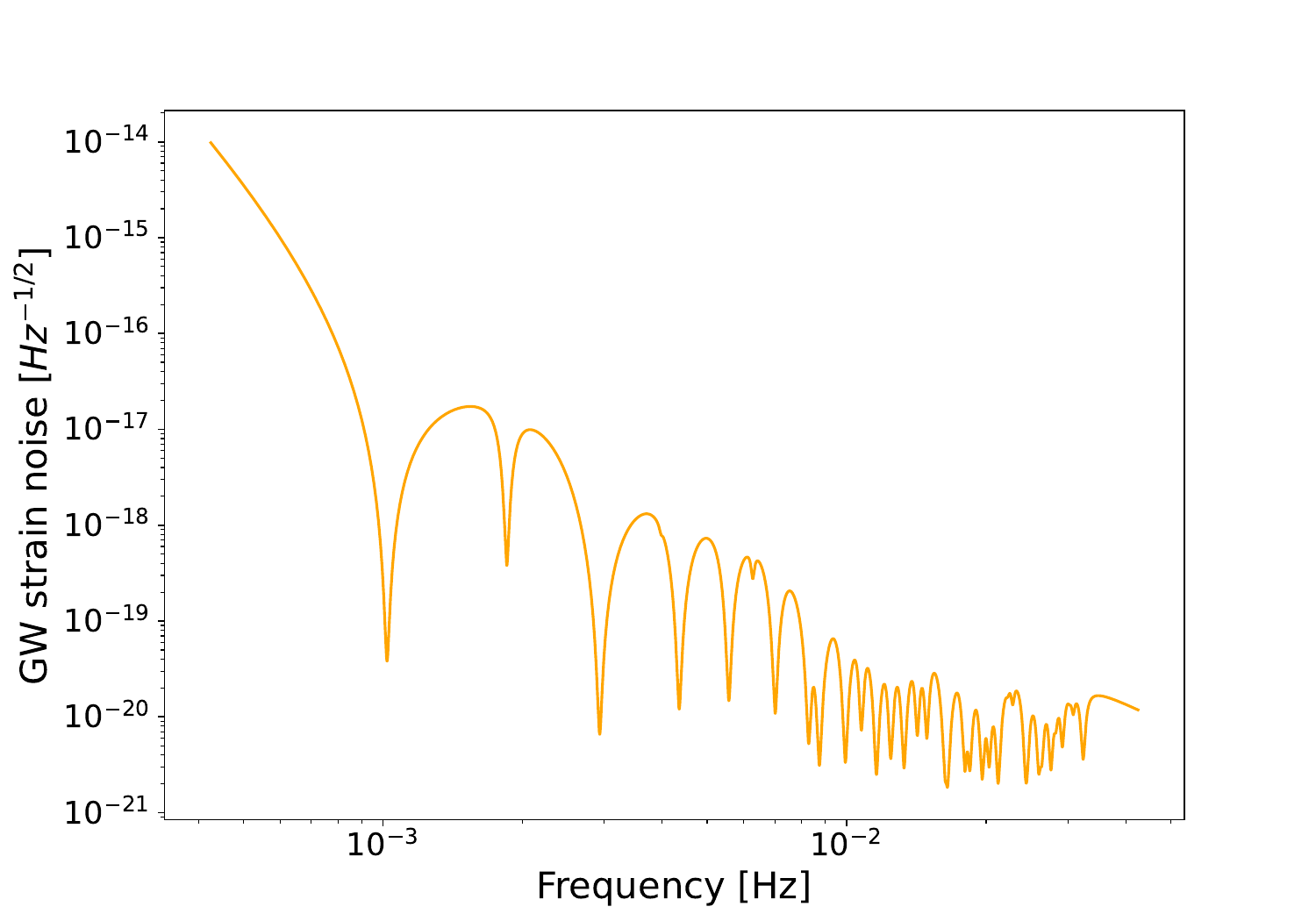}
\caption{Sensitivity of the LGWA experiment for the most recent detector concept ~\cite{van_Heijningen_2023} as function
 of frequency $\nu$ with $Q_n={\rm const}$. using the response function given in Eq.~(\ref{barg}).}
\label{fig:sensitivity_GreensOld}
\end{figure}

We observe a general trend of decreasing response at higher resonances,
its degree being dependent on the frequency dependence of the quality factor
$Q_n$.
At the same time, the distance between resonance frequencies decreases as we
move to higher frequencies in Fig.~\ref{fig:GWResponsePerUnitStrain_OldGreens}
indicating that the Moon becomes a broadband detector for $\omega\gg\omega_1$.
At frequencies smaller than the first eigenfrequency, we observe a strong
suppression of the response. This implies that the detectability of  GWs
with frequencies less than mHz is unlikely.

We can summarize our results in the  sensitivity plot shown in
Fig.~\ref{fig:sensitivity_GreensOld}. Using the minimal acceleration
predicted for two different detector concepts proposed for the LGWA experiment,
we show the  detection capabilities of the proposed LGWA experiment
for \(\tensor[_2]{S}{_2}\) modes in the Moon model~1. Using the values from
Table~\ref{table:MoonModelsCompareResponse}, the sensitivity curves
for other models can be obtained performing a simple overall-shift
of the curves for Model~1.

\begin{figure}[htp]
\centering
\includegraphics[width=0.49\textwidth]{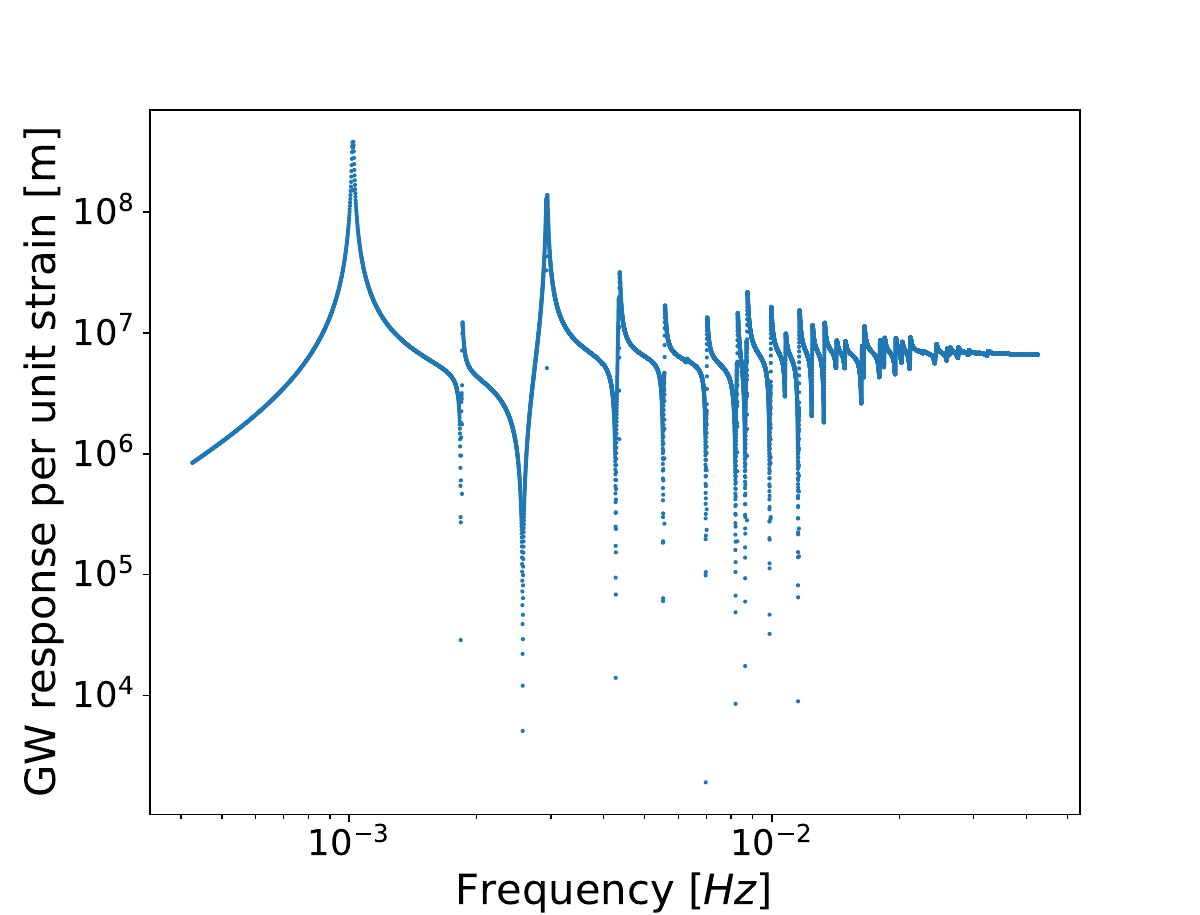}
\caption{The gravitational wave response in Moon model 1 including the first
  37 eigenfrequencies with  $Q_n={\rm const}$ using the response function given in Eq.~(\ref{eqn:Greens_Harms}).}
\label{fig:GWResponsePerUnitStrain_GreensHarms}
\end{figure}

Finally, we want to compare our results with earlier ones, in particular
with those shown by the LGWA collaboration in Fig.~1 from
Ref.~\cite{LGWA:2020mma}. In order to compare  more easily our results.
We choose now the same quality factor of $Q_n = 200$ for  all
eigenfrequencies. Moreover, we use now instead of
Eq.~(\ref{barg}) the Green function employed in Ref.~\cite{LGWA:2020mma},
\begin{equation}
\bar{g}(t) = \frac{1}{\omega_n^2-\omega_{0}^2 + \i\omega_n^2/Q_n} .
\label{eqn:Greens_Harms}
\end{equation}
The resulting gravitational response
per unit strain as a function of the frequency of the incoming GW is shown
in Fig.~\ref{fig:GWResponsePerUnitStrain_GreensHarms}. Compared to Fig.~1
Ref.~\cite{LGWA:2020mma}, we note that the positions of the first
resonance as well as the overall-shape agree well, while the asymptotic
value for large frequencies in our case is a factor few higher.

Note that while the response for our default Green function is real,
we have to take the real part of the complex response obtained using 
Eq.~(\ref{eqn:Greens_Harms}). Consequently, the strain in
Fig.~\ref{fig:GWResponsePerUnitStrain_GreensHarms} oscillates around the
asymptotic value, while the strain in
Fig.~\ref{fig:GWResponsePerUnitStrain_OldGreens} approaches the asymptotic
value from above. Taking into account this difference, our results for the
two different Green functions agree for large frequencies (and the same
value for the quality factor $Q_n$). On the other hand, the larger value of
quality factor $Q_n=3300$ used by us in the main part of our analysis
explains the larger sensitivities found by us. This difference in the
used quality factor explains also the difference in the overall scale seen
in the sensitivity plot~\ref{fig:GWResponsePerUnitStrain_GreensHarms}, while
the variation in shape is caused by the differences in the Green function
used.

\begin{figure}[htp]
\centering
\includegraphics[width=0.49\textwidth]{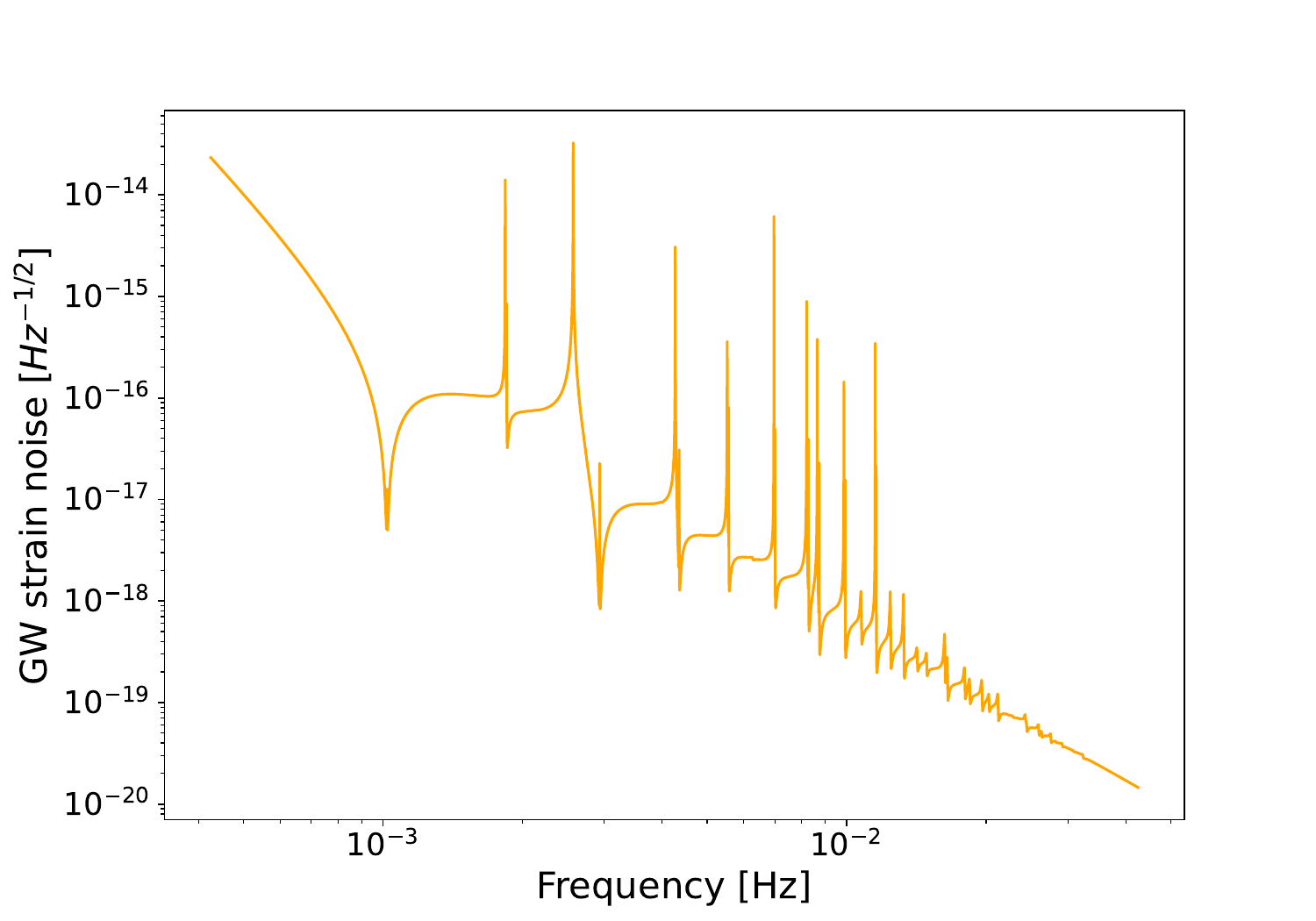}
\caption{Sensitivity of the LGWA experiment for the most recent detector concept ~\cite{van_Heijningen_2023} as function
 of frequency $\nu$ with $Q_n={\rm const}$ using the response function given in Eq.~(\ref{eqn:Greens_Harms}).}
\label{fig:sensitivity_GreensHarms}
\end{figure}

\section{Conclusions}

We have studied the response of the Moon to gravitational perturbations in
general scalar-tensor theories of gravity. Our semi-analytic study was
based on the approach developed by Alterman {\it et al.}~\cite{Alterman59}
for the study of seismic waves in the Earth. Its main limitation is the
restriction to heterogeneous, but spherically symmetric Moon models.

We have analyzed three sets of Moon models which are based on different
methodologies to determine the moon interior. The variation of these
models is largest in the first Lam\'e parameter $\lambda$ which determines
the response to bulk forces and therefore does not contribute to spheroidal
oscillations. As a result, the displacement
and the eigenfrequencies of the first eigenmodes determined by us
numerically agree relatively well for the three different Moon models;
the variations between the different models are however increasing for larger
$n$. Using the reach in measuring accelerations predicted for the LGWA
experiment from Ref.~\cite{LGWA:2020mma}, we found a nominal sensitivity
of this experiment to GWs with amplitude $h\simeq 10^{-20}$ for
quality factors $Q_n\simeq 3300$ in the mHz range.

\section*{Acknowledgements}

This work was motivated by the presentations at the ``Lunar session'' of the
2022 Vulcano Workshop ``Frontier Objects in Astrophysics and Particle
Physics''. We would like to thank the organizers and participants, and in
particular the late Stavros Katsanevas, for raising our interest
in this topic. We are grateful to Jan Harms for useful comments on the
manuscript and the LGWA experiment.

\appendix

\section{Reduction to first-order equations}
\label{reduction}

In this appendix, we will transform the Euler and Poisson equation to
a system of first-order equations, which is more suitable both
for numerical integration and for imposing
the appropriate boundary conditions, 
following the approach of Ref.~\cite{SeismicWaveAndSources}.

\subsection{Boundary conditions}

We have to impose the following four boundary conditions:
\begin{enumerate}
\item The solution is well defined at the origin.
\item The stresses vanish at the deformed surfaces and stay continuous at an internal deformed surface of discontinuity. 
\item The displacements are continuous at an internal surface of discontinuity, with the exception of a solid-liquid interface where only the radial displacement is continuous. 
\item The gravitational potential and its radial derivative are continuous at the deformed surface of the earth at an internal deformed surface of discontinuity. 
\end{enumerate}
In order to implement these boundary conditions mathematically,  we consider
the stresses close to a surface of discontinuity at \(r = c\)  in a strained
state,
\begin{equation}
\sigma_{jk}(c+u_r) = \sigma_{jk}(c) + u_r\left(\frac{\partial\sigma_{jk}}{\partial r}\right)_{r=c} = \sigma^{(e)}_{jk}(c) - P^{(0)}(c)\delta_{jk}.
\end{equation}
The additional elastic stresses at \(c\) and at \(c + u_r\) are equal to 
first order in \(u_r\). We also see that a small element of the medium
carries its initial stress with it when it moves from one place to another.
The boundary condition 4 regards only the gravitational potential.
Mathematically it says that
\begin{equation}
  \Psi_{<} = \Psi_{>} \quad\text{ and }\quad
  \frac{\d\Psi_{<}}{\d r} = \frac{\d\Psi_{>}}{\d r} \quad\text{at } r = c+u_r, 
\end{equation}
where the indices \(<\) and \(>\) indicate that \(\Psi\) and $\Psi'$ are
evaluated at opposite sides of the surface of discontinuity. Expanding
\(\Psi\) around its equilibrium value and using the Poisson equation,
it follows
\begin{equation}
  \frac{\d^2\Psi^{(0)}}{\d r^2} +\frac{2}{r}\frac{\d\Psi^{(0)}}{\d r} =
  -4\pi G\rho^{(0)}
\end{equation}
and
\begin{equation}
  \psi_{<} = \psi_{>}, \qquad
  \dot{\psi}_< - 4\pi G\rho_<^{(0)}u_r = \dot{\psi}_> - 4\pi G\rho_>^{(0)}u_r.
\end{equation}
At the surface of the Moon we then must have that,
\begin{equation}
    \psi = \psi^{(e)}, \quad\Dot{\psi}-4\pi G\rho^{(0)} u_r = \Dot{\psi}^{(e)},
\end{equation}
where \(\psi_e\) is the gravitational potential outside the spherical model.

\subsection{Toroidal Oscillations}

For purely toroidal oscillations \(u_r\) and \(u_{j,j}\) both vanish.
The Euler equation~\eqref{euler} then simplifies to
\begin{widetext}
\begin{equation}
  \mu \Delta u_{j} + \frac{\d\mu}{\d r}\left(2\frac{\partial u_j}{\partial r} +
    \epsilon_{j a b}\hat{e}^{(r)}_a\epsilon_{b c d} \nabla_c u_{d}\right) +
  \omega^2\rho^{(0)}u_j = 0.
    \label{eqn:_toroidal_diff_equation}
\end{equation}
Setting $U^{(n)}=V^{(n)}=0$ in Eq.~(\ref{eqn:displ_gen})  appropriate for
toroidal oscillations, we can write the displacement as
\begin{equation}
  u_j(\vec{r}) =  \sum_{\sigma,m,l} Q^{(nml)}_j(\vec{r}) =
  \sum_{\sigma, m, l} y_1(r)\sqrt{l(l+1)}C^{(\sigma ml)}_{j}(\theta, \phi),\quad\quad  \sigma = c, s.
    \label{eqn:toroidal_displacement}
\end{equation}
Here, we have split the displacement into a sum over the core and mantle
contributions, $\sigma=\{c,s\}$, to make sure that the solution is well
defined at both the origin and in all parts of the mantle. We insert our
ansatz into \eqref{euler} and arrive at a new differential
equation for \(y_1\),
\begin{equation}
  \mu\left(\frac{\d^2y_1}{\d r^2}+\frac{2}{r}\frac{\d y_1}{\d r}\right)+
  \frac{\d\mu}{\d r}\left(\frac{\d y_1}{\d r}-\frac{y_1}{r}\right) +
  \omega^2\rho^{(0)}y_1 -\frac{l(l+1)}{r^2}\mu y_1 = 0.
  \label{y1}
\end{equation}
By boundary condition 2 we must have that the stresses vanish at the
core-mantle boundary. We therefore define a function \(y_2\) such that this
function is zero at the boundary. We write the stress in the radial direction
and define \(y_2\) as
\begin{equation}
    \hat{e}^{(r)}_{i}\sigma_{ij} = \sum_{\sigma,m,l}y_2(r)\sqrt{l(l+1)}\vec{C}^{(\sigma ml)}_{j}(\theta,\phi), \quad\quad  \sigma = c, s.
\end{equation}
We find \(y_2\) more explicitly inserting
Eq.~\eqref{eqn:toroidal_displacement} into \(\sigma_{ij}(\vec{u})\),
\begin{equation}
    y_2 = \mu\left(\frac{dy_1}{dr}-\frac{y_1}{r}\right).
\end{equation}
Boundary condition 2 requires then the following conditions on \(y_2\),
\begin{equation}
    y_2 = 0 \quad\quad\quad \text{at}\quad r = r_c\quad \text{and}\quad r =  R.\label{eqn:toroidal_numerical_bc}
\end{equation}
In order to avoid numerical problems caused by the derivative $\d\mu/\d r$ in
Eq.~(\ref{y1}), it is more convenient to use the equivalent system of
differential equations
\begin{subequations}
\begin{equation}
    \frac{\d y_1}{\d r} = \frac{y_1}{r} + \frac{y_2}{\mu},
\end{equation}
\begin{equation}
    \frac{\d y_2}{\d r} = \left(\frac{l^2+l-2}{r^2}\mu - \omega^2\rho^{(0)}\right)y_1 - \frac{3}{r}y_2.
\end{equation}\label{eqn:toroidal_numericall_diff_eqs}
\end{subequations}

\subsection{Spheroidal Oscillations}

Spheroidal oscillations were defined as displacements involving the vector
surface harmonics \(\vec{P}\) and \(\vec{B}\) only. Thus a
spheroidal displacement can be written as
\begin{equation}
  u_j(\vec{r}) =
 \sum_{\sigma,m,l} Q^{(nml)}_j(\vec{r}) =
  \sum_{\sigma,m,l}\left(y_{1n}(r)P^{(\sigma ml)}_j(\theta,\phi) + y_{3n}(r)\sqrt{l(l+1)}B^{(\sigma ml)}_j(\theta,\phi)\right).
    \label{eqn:spheroidal_displacement_1b}
\end{equation}
We decompose the gravitational perturbation as
\begin{equation}
    \psi(\vec{r}) = \sum_{\sigma, m, l}y_{5n}(r)Y^{(\sigma) m}_l(\theta,\phi).
    \label{eqn:spheroidal_gravity_eq1} 
\end{equation}
Inserting Eqs.~\eqref{eqn:spheroidal_displacement_1} and
\eqref{eqn:spheroidal_gravity_eq1} into the Euler and Poisson equations
results in a system of differential
equations for \(y_1, y_3\) and \(y_5\).
A lengthy, but not too complicated calculation of inserting
Eq.~\eqref{eqn:spheroidal_displacement_1} into \eqref{euler} and setting the
coefficients of \(P^{(ml)}_i\) and \(B^{(ml)}_i\) to zero leads to the system
\begin{align}
    &\mu\left(2\frac{\d X}{\d r} - \frac{l(l+1)}{r}Z\right) + \frac{\d(\lambda X)}{\d r} + 2\Dot{\mu}\frac{\d y_1}{\d r}
    + \rho^{(0)}\left[\frac{\d y_5}{\d r}-4\pi G\rho^{(0)} y_1 + g^{(0)}\left(X-\frac{\d y_1}{\d r}+\frac{2}{r}y_1 + \omega^2y_1\right)\right] = 0,\\
    &(\lambda + 2\mu)\frac{X}{r} - \frac{\d}{\d r}(\mu Z) - \mu\frac{Z}{r} + 2\Dot{\mu}\left(\frac{\d y_3}{\d r}+Z\right)
    +\rho_0\left(\frac{1}{r}(y_5-g^{(0)}y_1)+\omega^2y_3\right) = 0
\end{align}
with
\be
    X = \frac{\d y_1}{\d r} + \frac{2}{r}y_1 -\frac{l(l+1)}{r}y_3, \und
    Z = \frac{1}{r}(y_1-y_3)-\frac{\d y_3}{\d r}.
\ee
To accommodate the boundary conditions and to obtain a system of
first-order differential equations, we evaluate the elastic stress.
Setting
\begin{equation}
    \hat{e}^{(r)}_i\sigma_{(e)}^{ij} = \sum_{\sigma,m,l}\left(y_{2}(r)P^{(\sigma ml)}_j(\theta,\phi) + y_{4}(r)\sqrt{l(l+1)}B^{(\sigma ml)}_j(\theta,\phi)\right),
\end{equation}
we find \(y_2\) and \(y_4\) as
\begin{align}
  y_2 &= \lambda X + 2\mu\frac{\d y_1}{\d r} =
        (\lambda+2\mu)\frac{\d y_1}{\d r} + \frac{2\lambda}{r}y_1 - \lambda\frac{l(l+1)}{r}y_3,\\
    y_4 &= \mu\left(Z+2\frac{\d y_3}{\d r}\right) = \mu\left(\frac{1}{r}(y_1-y_3)+\frac{\d y_3}{\d r}\right).
\end{align}
We now insert \eqref{eqn:spheroidal_gravity_eq1} into \eqref{poisson} to obtain a differential equation for \(y_5\), 
\begin{equation}
    \frac{\d^2y_5}{\d r^2} + \frac{2}{r}\frac{\d y_5}{\d r}-\frac{l(l+1)}{r^2}y_5 = 4\pi G(\rho^{(0)} X + \dot{\rho}^{(0)}y_1).
\end{equation}
With \(\Delta\psi_e = 0\) outside the boundary of the spherical model,
the boundary condition becomes
\begin{equation}
    \frac{\d y_5}{\d r} -4\pi G\rho^{(0)}y_1 = -\frac{l+1}{r}y_5\quad\text{at}\quad r=R. 
\end{equation}
Defining a new function \(y_6\) as
\begin{equation}
    y_6 = \frac{\d y_5}{\d r} - 4\pi G\rho^{(0)}y_1 ,
\end{equation}
the boundary conditions at $r=R$ become
\begin{equation}
    y_2 = 0, \quad\quad y_4 = 0, \quad\quad y_6 + \frac{l+1}{r}y_5 = 0 .
\end{equation}
We now treat \(y_1, y_2, ..., y_6\) as independent variables, obtaining
a system of first-order differential equations valid in the mantle,
\begin{subequations}
    \begin{equation}
        \frac{\d y_1}{\d r} = -\frac{2\lambda}{(\lambda +2\mu)r}y_1 + \frac{1}{\lambda+2\mu}y_2 + \frac{l(l+1)\lambda}{(\lambda+2\mu)r}y_3,
    \end{equation}
    \begin{align}
    \frac{\d y_2}{\d r} &= \left[-\omega^2\rho^{(0)} - 4\frac{g^{(0)}\rho_0}{r}+\frac{4\mu(3\lambda+2\mu)}{(\lambda+2\mu)r^2}\right]y_1 - \frac{4\mu}{(\lambda+2\mu)r}y_2 + \frac{l(l+1)}{r}\left[g^{(0)}\rho^{(0)} - \frac{2\mu(3\lambda+2\mu)}{(\lambda+2\mu)r}\right]y_3 + \frac{l(l+1)}{r}y_4 -\rho^{0)} y_6
    \end{align}
    \begin{equation}
        \frac{\d y_3}{\d r} = -\frac{1}{r}y_1 + \frac{1}{r}y_3 + \frac{1}{\mu}y_4
    \end{equation}
    \begin{align}
        \frac{\d y_4}{\d r} &= \left[\frac{g^{(0)}\rho^{(0)}}{r}-\frac{2\mu(3\lambda+2\mu)}{(\lambda+2\mu)r^2}\right]y_1 - \frac{\lambda}{(\lambda+2\mu)r}y_2\\
   & +\left(-\omega^2\rho^{(0)} + ((2l^2+2l-1)\lambda + 2(l^2+l-1)\mu)\frac{2\mu}{(\lambda+2\mu)r^2}\right)y_3
    -\frac{3}{r}y_4 -\frac{\rho_0}{r}y_5\label{eqn:chapter4_differential_y4}
    \end{align}
    \begin{equation}
        \frac{\d y_5}{\d r} = 4\pi G\rho^{(0)} y_1 +y_6,
    \end{equation}
    \begin{equation}
        \frac{\d y_6}{\d r} = -4\pi\frac{l(l+1)}{r}G\rho^{(0)} y_3 + \frac{l(l+1)}{r^2}y_5 - \frac{2}{r}y_6.
    \end{equation}\label{eqn:spheroidal_numerical_diff_eqns_mantle}
\end{subequations}
In the core on the other hand, we have
\begin{equation}
    \mu = 0,\quad\quad y_2 = \lambda X,\quad\quad y_4 = 0,
    \label{eqn:core_restrictions}
\end{equation}
resulting in a simpler system of differential equations,
\begin{subequations}
    \begin{equation}
    \frac{\d y_1}{\d r} = -\frac{2}{r}y_1+\frac{1}{\lambda}y_2 + \frac{l(l+1)}{r}y_3,
    \end{equation}
    \begin{equation}
        \frac{\d y_2}{\d r} = -\left(\omega^2\rho^{(0)} + \frac{4g^{(0)}\rho^{(0)}}{r}\right)y_1 + \frac{l(l+1)}{r}g^{(0)}\rho^{(0)}y_3-\rho^{(0)}y_6,
    \end{equation}
    \begin{equation}
        \frac{\d y_5}{\d r} =4\pi G\rho^{(0)} y_1 + y_6
    \end{equation}
    \begin{equation}
        \frac{\d y_6}{\d r} =-4\pi\frac{l(l+1)}{r}G\rho^{(0)}y_3 + \frac{l(l+1)}{r^2}y_5-\frac{2}{r}y_6.
    \end{equation}\label{eqn:spheroidal_numerical_diff_eqns_core}
\end{subequations}
The parameter function \(y_4\) is zero in the core. We find an expression
for \(y_3\) using \eqref{eqn:chapter4_differential_y4} and
\eqref{eqn:core_restrictions},
\begin{equation}
    y_3 = \frac{1}{\omega^2r}\left(g_0y_1-\frac{1}{\rho_0}y_2-y_5\right).
\end{equation}

\section{Expansion of the Fourier transforms $\widetilde V^{(lm)}_{ij}$}

In this appendix, we will express the Fourier transforms
$\widetilde V^{(lm)}_{ij}(x)$ of the three vector surface harmonics in
terms of spherical Bessel functions $j_l(x)$ with Wigner's $6j$ symbols as
expansion coefficients. For the later, we use the following convention
\begin{align}
    \begin{pmatrix}
        l_1 & l_2 & l_3\\
        m_1 & m_2 & m_3
    \end{pmatrix}=&  W\sum_n\frac{(-1)^{l_1-l_2-m_3+n}}{n!}
    \left(\frac{\Pi_{i=1}^3(l_i+m_i)!(l_i-m_i)!}{(l_1+l_2-l_3-n)!(l_1-m_1-n)!(l_2+m_2-n)!}\right) \nn\\
    &\times(l_3-l_2+m_1+n)!(l_3-l_1-m_2+n)!,
\label{Wigner_def}
\end{align} 
with
\begin{equation}
    W = \sqrt{\left(\frac{\Pi_{i=1}^3(2p-2l_i)!}{(2p+1)!}\right)}\delta_{m_1+m_2-m_3}
\end{equation}
and \(2p = l_1 + l_2 + l_3\). The sum goes over positive values of \(n\) until
the denominator of the expression becomes negative.
The symbols are non-zero only, when 
\begin{equation}
    m_1 + m_2 + m_3 = 0,  
\text{ and }
  |l_a-l_b| \leq l_c \leq |l_a+l_b|, 
  \label{eqn:appendix_triangle_inequality}
\end{equation}
with $a, b, c = \{1, 2, 3\}$.
These two relations allow one to quickly determine if the Wigner symbols
are zero.

Our aim is to evaluate the Fourier transforms $V^{(ml)}_{ij}(kr)$ defined
by Eq.~(\ref{FourierV}) for the three cases
$\vec V^{(ml)}=\{\vec C^{(ml)}, \vec P^{(ml)},\vec B^{(ml)}\}$.
We begin with the $\vec C$ function, writing it in spherical
coordinates,
\begin{equation}
    \sqrt{l(l+1)}C^{(ml)}_j = \left(\hat{e}^{(\theta)}_j\frac{1}{ \sin{\theta}}\partial_\phi-\hat{e}^{(\phi)}_j\partial_\theta\right)Y^{(m)}_{(l)}.
\end{equation}
In order to unclutter the notation, we will omit the parentheses
around $m$ and $l$ in the spherical harmonics \(Y^m_l\), which we
define as
\begin{equation}
  Y^m_l(\theta,\phi) = \sqrt{\frac{(l-m)!}{(l+m)!}}P_l^m(\cos{\theta})\e^{\i m\phi}. \label{spherHarm}
\end{equation}
We can write this vector in Cartesian components as
\begin{equation}
  \sqrt{l(l+1)}C^{(ml)}_j =
  a^{(0)}_j Y^m_l + a^{(+)}_j Y^{m+1}_l + a^{(-)}_j Y^{m-1}_l,
    \label{eqn:appendix_C_expansion}
\end{equation}
where \( a^{(0)}_j\), \(a^{(+)}_j\) and \(a^{(-)}_j\) are combinations of
the Cartesian unit vectors given by
\begin{align}
    a^{(0)}_j &= -\i m\hat{e}_j^{(z)},\\
    a^{(\pm)}_j &=
    \frac{\i}{2}\sqrt{(l\mp m)(l\pm m+1)}(\hat{e}_j^{(x)}\mp \i\hat{e}_j^{(y)}) .
\end{align}
Assuming a general momentum vector,
\begin{equation}
    k_i = \omega(\sin{\psi}\cos{\chi},\sin{\psi}\sin{\chi},\cos{\psi}),
\end{equation}
we perform next a partial-wave expansion of the exponential,
\begin{equation}
  \e^{-\i\vec k\vec r} = \sum_{l_1 = 0}^\infty\sum_{m_1=-l_1}^{l_1}(2l_1+1)\i^{-l_1}
  j_{l_1}(k r)Y_{l_1}^{m_1}(\theta,\phi)Y_{l_1}^{*m_1}(\psi,\chi) ,
    \label{eqn:appendix_exponential_expansion}
\end{equation}
and inserting \eqref{eqn:appendix_C_expansion} and \eqref{eqn:appendix_exponential_expansion} into $\widetilde C^{(lm)}_{ij}(kr)$, we obtain
\begin{equation}
  \sqrt{l(l+1)} \widetilde C^{(ml)}_{jk}(kr) = 
   \sum_{l_1=0}^\infty\sum_{m_1=-l_1}^{l_1}(2l_1+1)
  \i^{-l_1}j_{l_1}(k r)Y_{l_1}^{*m_1}(\psi,\chi)
  \left(A^{(0)}_j a^{(0)}_k + A^{(+)}_j a^{(+)}_k + A^{(-)}_j a^{(-)}_k\right)
    \label{eqn:appendix_C_Integral_2}
\end{equation}
with
\be
A^{(n)}_j = \int_0^{2\pi}\int_0^\pi\hat{e}^{(r)}_\alpha
Y_{l_1}^{*m_1}(\theta,\phi)Y_{l}^{m+n}(\psi,\chi)
 \sin{\theta}d\theta d\phi .
\ee
Using the identity
\begin{equation}
  \int_0^{2\pi}\d\phi\int_0^\pi \d\theta\sin\theta \:
  Y_{l_1}^{m_1}Y_{l_2}^{m_2}Y_{l_3}^{m_3}
  = 4\pi\begin{pmatrix}l_1&l_2&l_3\\ 0 & 0 & 0\end{pmatrix}\begin{pmatrix}l_1&l_2&l_3\\ m_1 & m_2 & m_3\end{pmatrix}
    \label{eqn:appendix_wigner_identity}
\end{equation}
in \eqref{eqn:appendix_C_Integral_2}, we arrive at the result
\begin{align}
   \sqrt{l(l+1)}\widetilde C^{(ml)}_{jk} & =
  4\pi\sum_{l_1=0}^\infty\sum_{m_1=-l_1}^{l_1}(2l_1+1)
  \i^{-l_1}j_{l_1}(kr)Y^{*m_1}_{l_1}(\psi,\chi)\begin{pmatrix}
        l_1 & l_2 & 1\\
        0 & 0 & 0
  \end{pmatrix}
  \sum_{k=-2}^2 A^{(k)}T^{(k)}
  \label{eqn:appendix_toroidal_integral_expansion}
\end{align}
%
with
%
%
%
\begin{equation}
        T^{(0)} = \begin{pmatrix}
            0 & 0 & 0\\
            0 & 0 & 0\\
            0 & 0 & \i
        \end{pmatrix}, \quad\quad T^{(\pm 1)} = \frac12\begin{pmatrix}
            0 & 0 & \mp\i\\
            0 & 0 & 1\\
            \mp\i & 1 & 0
        \end{pmatrix}, \quad\quad T^{(\pm 2)} = \mp \begin{pmatrix}
            \mp\i & 1 & 0\\
            1 & \pm\i & 0\\
            0 & 0 & 0
            \end{pmatrix},
\end{equation}
and
\begin{subequations}
    \begin{align}
        A^{(0)} = -m&\begin{pmatrix}
            l_1 & l & 1\\
            m_1 & m & 0
        \end{pmatrix}  
        -\frac{1}{2\sqrt{2}}\sqrt{(l+m)(l-m+1)}\begin{pmatrix}
            l_1 & l & 1\\
            m_1 & m-1 & 1
        \end{pmatrix}  \\
        &+\frac{1}{2\sqrt{2}}\sqrt{(l-m)(l+m+1)}\begin{pmatrix}
            l_1 & l & 1\\
            m_1 & m +1 & -1
        \end{pmatrix},
    \end{align}
    \begin{align}
        A^{(\pm 1)} = -\sqrt{2}m \begin{pmatrix}
            l_1 & l & 1\\
            m_1 & m\mp 1 & \mp 1
        \end{pmatrix}
        \mp \sqrt{(l\pm m)(l\mp m+1)}\begin{pmatrix}
            l_1 & l & 1\\
            m_1 & m\mp 1 & 0
        \end{pmatrix},
    \end{align}
    \begin{align}
      A^{(\pm 2)} &= \frac{1}{2\sqrt{2}}\sqrt{(l\pm m)(l\mp m+1)}\begin{pmatrix}
            l_1 & l & 1\\
            m_1 & m\mp 1 & -1
        \end{pmatrix},
    \end{align}
\end{subequations}
%


Next we consider the case of the $\vec P$ integral. Using that in spherical
coordinates
$P^{(ml)}_j = \hat{e}^{(r)}_j Y_l^m(\theta,\phi)$, we apply again the partial-wave
expansion \eqref{eqn:appendix_exponential_expansion}. The resulting
product \(\hat{e}^{r}_j\hat{e}^{r}_k\) can be written in terms of
Cartesian unit coordinates and the associated Legendre polynomial as
%
%
\begin{align}
    \hat{e}^{(r)}_j\hat{e}^{(r)}_k =& \frac{1}{3}\delta_{jk} + \frac{1}{3}Y^0_2(\hat{e}^{(x)}_j\hat{e}^{(x)}_k - \hat{e}^{(y)}_j\hat{e}^{(y)}_k + 2\hat{e}^{(z)}_j\hat{e}^{(z)}_k)
    + \frac{1}{\sqrt{6}}Y^1_2\left(\hat{e}^{(z)}_j\hat{e}^{(-)}_k + \hat{e}^{(z)}_k\hat{e}^{(-)}_j\right)\nn\\
    &-\frac{1}{\sqrt{6}}Y_2^{-1}\left(\hat{e}^{(z)}_j\hat{e}^{(+)}_k + \hat{e}^{(z)}_k\hat{e}^{(+)}_j\right)
     + \frac{1}{\sqrt{6}}Y_2^2\hat{e}^{(-)}_j\hat{e}^{(-)}_k + \frac{1}{\sqrt{6}}Y_2^{-2}\hat{e}^{(+)}_j\hat{e}^{(+)}_k, 
\end{align}
where we have defined $\hat{e}^{(\pm)} = \hat{e}^{(x)}\pm\i\hat{e}^{(y)}$. Then we employ the identity \(\eqref{eqn:appendix_wigner_identity}\) to obtain
\begin{align}
   \widetilde P^{(lm)}_{ij}(kr) = &
  4\pi\sum_{l_1=0}^\infty\sum_{m_1=-l_1}^{l_1}(2l_1+1)\i^{-l_1}j_{l_1}(k r)
  Y^{*m_1}_{l_1}(\psi, \chi)
    \begin{pmatrix}
        l_1 & l & 2\\
        0 & 0 & 0 
    \end{pmatrix} 
    \sum_{j=-2}^{2}\Gamma_{ij}^{(j)}\begin{pmatrix}
        l_1 & l & 2\\
        m_1 & m & j
    \end{pmatrix}
    \label{fourierP}
\end{align}
where the \(\Gamma^{(j)}\) symbols are defined as combinations of the
Cartesian unit vectors, 
%
\begin{align}
    \Gamma_{jk}^{(0)} &= \frac{1}{3}(-\hat{e}^{(x)}_j\hat{e}^{(x)}_k -\hat{e}^{(y)}_j\hat{e}^{(y)}_k + 2\hat{e}^{(z)}_j\hat{e}^{(z)}_k),\quad
    \Gamma_{jk}^{(\pm 1)} =\frac{1}{\sqrt{6}}
                            \left(\hat{e}^{(z)}_j\hat{e}^{(\mp)}_k +\hat{e}^{(\mp)}_j\hat{e}^{(z)}_k\right),
                            \quad
    \Gamma_{jk}^{(\pm 2)} = \frac{1}{\sqrt{6}}\hat{e}^{(\mp)}_j\hat{e}^{(\mp)}_k .
\end{align}

Finally we have to evaluate the Fourier transform of the vector
surface harmonics  $\vec B$. Inserting into
\begin{equation}
\sqrt{l(l+1)}B^{(ml)}_j = \epsilon_{ikj}\hat{e}^{(r)}_i C^{(ml)}_k
\end{equation}
the expansion \eqref{eqn:appendix_C_expansion} for \(C^{(ml)}_k\),
we obtain
\be
\sqrt{l(l+1)}\hat{e}^{(r)}_i B^{(ml)}_j =
\hat{e}^{(r)}_i(\epsilon_{abj}\hat{e}^{(r)}_a a^{(0)}_b)Y^m_l + \hat{e}^{(r)}_i(\epsilon_{abj}\hat{e}^{(r)}_a a^{(+)}_b)Y^{m+1}_l
    + \hat{e}^{(r)}_i(\epsilon_{abj}\hat{e}^{(r)}_a a^{(-)}_b)Y^{m-1}_l .
\ee
Next we represent the products
\(\hat{e}^{(r)}_i\epsilon_{abj}\hat{e}^{(r)}_a a^{(0,\pm)}_b\)
by the \(\Gamma\) symbols defined previously, dropping all antisymmetric
parts, obtaining
\begin{align}
  \label{fourierB}
  \sqrt{l(l+1)} \widetilde B^{(ml)}_{jk}(kr)
  = 4\pi\sum_{l_1=0}^\infty\sum_{m_1=-l_1}^{l_1}(2l_1+1)\i^{-l_1}&j_{l_1}(kr)Y_{l_1}^{*m_1}(\psi,\chi)\begin{pmatrix}
        l_1 & l & 0\\
        0 & 0 & 0
    \end{pmatrix}
    \sum_{j=-2}^{2}D^{(j)}\Gamma^{(j)}_{ik} .
\end{align}
where the \(D^{(j)}\) are given by
\begin{align*}
    D^{(0)} &=-\frac{3}{2\sqrt{6}}\sqrt{(l-m)(l+m+1)}\begin{pmatrix}
        l_1 & l & 2\\
        m_1 & m+1 & -1
    \end{pmatrix}
    -\frac{3}{2\sqrt{6}}\sqrt{(l+m)(l-m+1)}\begin{pmatrix}
        l_1 & l & 2\\
        m_1 & m-1 & 1
    \end{pmatrix},\\
    D^{(\pm 1)} &=-\frac{m}{2}\begin{pmatrix}
        l_1 & l & 2\\
        m_1 & m & \pm 1
    \end{pmatrix} - \frac{3}{2\sqrt{6}}\sqrt{(l\mp m)(l\pm m+1)}\begin{pmatrix}
        l_1 & l & 2\\
        m_1 & m\pm 1 & 0
    \end{pmatrix}
    -\frac{1}{2}\sqrt{(l\pm m)(l\mp m+1)}\begin{pmatrix}
        l_1 & l & 2\\
        m_1 & m\mp 1 & \pm 2
    \end{pmatrix},\\
    D^{(\pm 2)} &= \mp m\begin{pmatrix}
        l_1 & l & 2\\
        m_1 & m & \pm 2
    \end{pmatrix} -\frac{1}{2}\sqrt{(l\mp m)(l\pm m+1)}\begin{pmatrix}
        l_1 & l & 2\\
        m_1 & m\pm 1 & \pm 1
    \end{pmatrix} .
\end{align*}


\end{widetext}


\begin{thebibliography}{26}
\expandafter\ifx\csname natexlab\endcsname\relax\def\natexlab#1{#1}\fi
\expandafter\ifx\csname bibnamefont\endcsname\relax
  \def\bibnamefont#1{#1}\fi
\expandafter\ifx\csname bibfnamefont\endcsname\relax
  \def\bibfnamefont#1{#1}\fi
\expandafter\ifx\csname citenamefont\endcsname\relax
  \def\citenamefont#1{#1}\fi
\expandafter\ifx\csname url\endcsname\relax
  \def\url#1{\texttt{#1}}\fi
\expandafter\ifx\csname urlprefix\endcsname\relax\def\urlprefix{URL }\fi
\providecommand{\bibinfo}[2]{#2}
\providecommand{\eprint}[2][]{\url{#2}}

\bibitem[{\citenamefont{Weber}(1960)}]{WeberPhysRev.117.306}
\bibinfo{author}{\bibfnamefont{J.}~\bibnamefont{Weber}},
  \bibinfo{journal}{Phys. Rev.} \textbf{\bibinfo{volume}{117}},
  \bibinfo{pages}{306} (\bibinfo{year}{1960}),
  \urlprefix\url{https://link.aps.org/doi/10.1103/PhysRev.117.306}.

\bibitem[{\citenamefont{Dyson}(1969)}]{Dyson:1969zgf}
\bibinfo{author}{\bibfnamefont{F.~J.} \bibnamefont{Dyson}},
  \bibinfo{journal}{Astrophys. J.} \textbf{\bibinfo{volume}{156}},
  \bibinfo{pages}{529} (\bibinfo{year}{1969}).

\bibitem[{\citenamefont{Ben-Menahem}(1983)}]{Ben-Menahem:1983}
\bibinfo{author}{\bibfnamefont{A.}~\bibnamefont{Ben-Menahem}},
  \bibinfo{journal}{Il Nuovo Cimento} \textbf{\bibinfo{volume}{C6}},
  \bibinfo{pages}{49} (\bibinfo{year}{1983}).

\bibitem[{\citenamefont{Alterman et~al.}(1959)\citenamefont{Alterman, Jarosch,
  and Pekeris}}]{Alterman59}
\bibinfo{author}{\bibfnamefont{Z.}~\bibnamefont{Alterman}},
  \bibinfo{author}{\bibfnamefont{H.}~\bibnamefont{Jarosch}}, \bibnamefont{and}
  \bibinfo{author}{\bibfnamefont{C.~L.} \bibnamefont{Pekeris}},
  \bibinfo{journal}{Proc. Roy. Soc.} \textbf{\bibinfo{volume}{A252}},
  \bibinfo{pages}{80} (\bibinfo{year}{1959}).

\bibitem[{\citenamefont{Wiggins and
  Press}(1969)}]{https://doi.org/10.1029/JB074i022p05351}
\bibinfo{author}{\bibfnamefont{R.~A.} \bibnamefont{Wiggins}} \bibnamefont{and}
  \bibinfo{author}{\bibfnamefont{F.}~\bibnamefont{Press}},
  \bibinfo{journal}{Journal of Geophysical Research (1896-1977)}
  \textbf{\bibinfo{volume}{74}}, \bibinfo{pages}{5351} (\bibinfo{year}{1969}),
  \urlprefix\url{https://agupubs.onlinelibrary.wiley.com/doi/abs/10.1029/JB074i022p05351}.

\bibitem[{\citenamefont{{Mast} et~al.}(1974)\citenamefont{{Mast}, {Nelson}, and
  {Saarloos}}}]{1974ApJ...187L..49M}
\bibinfo{author}{\bibfnamefont{T.~S.} \bibnamefont{{Mast}}},
  \bibinfo{author}{\bibfnamefont{J.~E.} \bibnamefont{{Nelson}}},
  \bibnamefont{and} \bibinfo{author}{\bibfnamefont{J.~A.}
  \bibnamefont{{Saarloos}}}, \bibinfo{journal}{Astrophys.\ J.\ Lett.}
  \textbf{\bibinfo{volume}{187}}, \bibinfo{pages}{L49} (\bibinfo{year}{1974}).

\bibitem[{\citenamefont{Coughlin and Harms}(2014)}]{Coughlin:2014xua}
\bibinfo{author}{\bibfnamefont{M.}~\bibnamefont{Coughlin}} \bibnamefont{and}
  \bibinfo{author}{\bibfnamefont{J.}~\bibnamefont{Harms}},
  \bibinfo{journal}{Phys. Rev. D} \textbf{\bibinfo{volume}{90}},
  \bibinfo{pages}{042005} (\bibinfo{year}{2014}), \eprint{1406.1147}.

\bibitem[{\citenamefont{Majstorovi\'c et~al.}(2019)\citenamefont{Majstorovi\'c,
  Rosat, and Rogister}}]{Majstorovic:2019fog}
\bibinfo{author}{\bibfnamefont{J.}~\bibnamefont{Majstorovi\'c}},
  \bibinfo{author}{\bibfnamefont{S.}~\bibnamefont{Rosat}}, \bibnamefont{and}
  \bibinfo{author}{\bibfnamefont{Y.}~\bibnamefont{Rogister}},
  \bibinfo{journal}{Phys. Rev. D} \textbf{\bibinfo{volume}{100}},
  \bibinfo{pages}{044048} (\bibinfo{year}{2019}), \bibinfo{note}{[Erratum:
  Phys.Rev.D 103, 029901 (2021)]}.

\bibitem[{\citenamefont{Amaro-Seoane et~al.}(2021)\citenamefont{Amaro-Seoane,
  Bischof, Carter, Hartig, and Wilken}}]{Amaro-Seoane:2020ahu}
\bibinfo{author}{\bibfnamefont{P.}~\bibnamefont{Amaro-Seoane}},
  \bibinfo{author}{\bibfnamefont{L.}~\bibnamefont{Bischof}},
  \bibinfo{author}{\bibfnamefont{J.~J.} \bibnamefont{Carter}},
  \bibinfo{author}{\bibfnamefont{M.-S.} \bibnamefont{Hartig}},
  \bibnamefont{and} \bibinfo{author}{\bibfnamefont{D.}~\bibnamefont{Wilken}},
  \bibinfo{journal}{Class. Quant. Grav.} \textbf{\bibinfo{volume}{38}},
  \bibinfo{pages}{125008} (\bibinfo{year}{2021}), \eprint{2012.10443}.

\bibitem[{\citenamefont{Jani and Loeb}(2020)}]{Jani:2020gnz}
\bibinfo{author}{\bibfnamefont{K.}~\bibnamefont{Jani}} \bibnamefont{and}
  \bibinfo{author}{\bibfnamefont{A.}~\bibnamefont{Loeb}}
  (\bibinfo{year}{2020}), \eprint{2007.08550}.

\bibitem[{\citenamefont{Harms et~al.}(2021)}]{LGWA:2020mma}
\bibinfo{author}{\bibfnamefont{J.}~\bibnamefont{Harms}} \bibnamefont{et~al.}
  (\bibinfo{collaboration}{LGWA}), \bibinfo{journal}{Astrophys. J.}
  \textbf{\bibinfo{volume}{910}}, \bibinfo{pages}{1} (\bibinfo{year}{2021}),
  \eprint{2010.13726}.

\bibitem[{\citenamefont{Katsanevas and Bernard}(2020)}]{LSGA}
\bibinfo{author}{\bibfnamefont{S.}~\bibnamefont{Katsanevas}} \bibnamefont{and}
  \bibinfo{author}{\bibfnamefont{P.}~\bibnamefont{Bernard}},
  \emph{\bibinfo{title}{Lunar Seismic and Gravitational Antenna, in Ideas for
  exploring the Moon with a large European lander}} (\bibinfo{publisher}{ESA},
  \bibinfo{year}{2020}).

\bibitem[{\citenamefont{Amaro-Seoane et~al.}(2017)}]{LISA:2017pwj}
\bibinfo{author}{\bibfnamefont{P.}~\bibnamefont{Amaro-Seoane}}
  \bibnamefont{et~al.} (\bibinfo{collaboration}{LISA}) (\bibinfo{year}{2017}),
  \eprint{1702.00786}.

\bibitem[{\citenamefont{{ET Steering Committee}}(2020)}]{ET}
\bibinfo{author}{\bibnamefont{{ET Steering Committee}}},
  \emph{\bibinfo{title}{{Einstein Telescope design report update, available
  from European Gravitational Observatory, document number ET-0007B-20. }}}
  (\bibinfo{year}{2020}).

\bibitem[{\citenamefont{James R.~Bates and Kernaghan}(1979)}]{ALSEP}
\bibinfo{author}{\bibfnamefont{W.~L.} \bibnamefont{James R.~Bates}}
  \bibnamefont{and}
  \bibinfo{author}{\bibfnamefont{H.}~\bibnamefont{Kernaghan}},
  \emph{\bibinfo{title}{{ALSEP Termination Report}}} (\bibinfo{publisher}{NASA
  Reference Publication 1036}, \bibinfo{year}{1979}).

\bibitem[{\citenamefont{Kupfer et~al.}(2018)\citenamefont{Kupfer, Korol, Shah,
  Nelemans, Marsh, Ramsay, Groot, Steeghs, and Rossi}}]{Kupfer:2018jee}
\bibinfo{author}{\bibfnamefont{T.}~\bibnamefont{Kupfer}},
  \bibinfo{author}{\bibfnamefont{V.}~\bibnamefont{Korol}},
  \bibinfo{author}{\bibfnamefont{S.}~\bibnamefont{Shah}},
  \bibinfo{author}{\bibfnamefont{G.}~\bibnamefont{Nelemans}},
  \bibinfo{author}{\bibfnamefont{T.~R.} \bibnamefont{Marsh}},
  \bibinfo{author}{\bibfnamefont{G.}~\bibnamefont{Ramsay}},
  \bibinfo{author}{\bibfnamefont{P.~J.} \bibnamefont{Groot}},
  \bibinfo{author}{\bibfnamefont{D.~T.~H.} \bibnamefont{Steeghs}},
  \bibnamefont{and} \bibinfo{author}{\bibfnamefont{E.~M.} \bibnamefont{Rossi}},
  \bibinfo{journal}{Mon. Not. Roy. Astron. Soc.}
  \textbf{\bibinfo{volume}{480}}, \bibinfo{pages}{302} (\bibinfo{year}{2018}),
  \eprint{1805.00482}.

\bibitem[{\citenamefont{Jordan}(1959)}]{Jordan59}
\bibinfo{author}{\bibfnamefont{P.}~\bibnamefont{Jordan}}, \bibinfo{journal}{Z.
  Phys.} \textbf{\bibinfo{volume}{157}}, \bibinfo{pages}{112}
  (\bibinfo{year}{1959}).

\bibitem[{\citenamefont{Brans and Dicke}(1961)}]{Brans:1961sx}
\bibinfo{author}{\bibfnamefont{C.}~\bibnamefont{Brans}} \bibnamefont{and}
  \bibinfo{author}{\bibfnamefont{R.~H.} \bibnamefont{Dicke}},
  \bibinfo{journal}{Phys. Rev.} \textbf{\bibinfo{volume}{124}},
  \bibinfo{pages}{925} (\bibinfo{year}{1961}).

\bibitem[{\citenamefont{van Heijningen et~al.}(2023)\citenamefont{van
  Heijningen, ter Brake, Gerberding, Chalathadka~Subrahmanya, Harms, Bian,
  Gatti, Zeoli, Bertolini, Collette et~al.}}]{van_Heijningen_2023}
\bibinfo{author}{\bibfnamefont{J.~V.} \bibnamefont{van Heijningen}},
  \bibinfo{author}{\bibfnamefont{H.~J.~M.} \bibnamefont{ter Brake}},
  \bibinfo{author}{\bibfnamefont{O.}~\bibnamefont{Gerberding}},
  \bibinfo{author}{\bibfnamefont{S.}~\bibnamefont{Chalathadka~Subrahmanya}},
  \bibinfo{author}{\bibfnamefont{J.}~\bibnamefont{Harms}},
  \bibinfo{author}{\bibfnamefont{X.}~\bibnamefont{Bian}},
  \bibinfo{author}{\bibfnamefont{A.}~\bibnamefont{Gatti}},
  \bibinfo{author}{\bibfnamefont{M.}~\bibnamefont{Zeoli}},
  \bibinfo{author}{\bibfnamefont{A.}~\bibnamefont{Bertolini}},
  \bibinfo{author}{\bibfnamefont{C.}~\bibnamefont{Collette}},
  \bibnamefont{et~al.}, \bibinfo{journal}{Journal of Applied Physics}
  \textbf{\bibinfo{volume}{133}} (\bibinfo{year}{2023}), ISSN
  \bibinfo{issn}{1089-7550},
  \urlprefix\url{http://dx.doi.org/10.1063/5.0144687}.

\bibitem[{\citenamefont{Poisson and Will}(2014)}]{Poisson14}
\bibinfo{author}{\bibfnamefont{E.}~\bibnamefont{Poisson}} \bibnamefont{and}
  \bibinfo{author}{\bibfnamefont{C.~M.} \bibnamefont{Will}},
  \emph{\bibinfo{title}{Gravity: Newtonian, Post-Newtonian, Relativistic}}
  (\bibinfo{publisher}{Cambridge University Press}, \bibinfo{year}{2014}).

\bibitem[{\citenamefont{Ben-Menahem and Singh}(1981)}]{SeismicWaveAndSources}
\bibinfo{author}{\bibfnamefont{A.}~\bibnamefont{Ben-Menahem}} \bibnamefont{and}
  \bibinfo{author}{\bibfnamefont{S.~J.} \bibnamefont{Singh}},
  \emph{\bibinfo{title}{Seismic Waves and Sources}}
  (\bibinfo{publisher}{Springer-Verlag Berlin Heidelberg New York},
  \bibinfo{year}{1981}).

\bibitem[{\citenamefont{{N{\"o}dtvedt, M. P.}}(2023)}]{Noedtvedt23}
\bibinfo{author}{\bibnamefont{{N{\"o}dtvedt, M. P.}}}, Master's thesis,
  \bibinfo{school}{{NTNU Trondheim, available at
  https://hdl.handle.net/11250/3080886}}, \bibinfo{address}{Trondheim}
  (\bibinfo{year}{2023}),
  \urlprefix\url{{https://hdl.handle.net/11250/3080886}}.

\bibitem[{\citenamefont{Garcia~Raphael et~al.}(2019)}]{GarciaLS:2019}
\bibinfo{author}{\bibfnamefont{F.}~\bibnamefont{Garcia~Raphael}}
  \bibnamefont{et~al.}, \bibinfo{journal}{Space Science Reviews}
  \textbf{\bibinfo{volume}{215}} (\bibinfo{year}{2019}).

\bibitem[{\citenamefont{Verma}(2021)}]{Verma:2021nbz}
\bibinfo{author}{\bibfnamefont{P.}~\bibnamefont{Verma}},
  \bibinfo{journal}{Universe} \textbf{\bibinfo{volume}{7}},
  \bibinfo{pages}{235} (\bibinfo{year}{2021}).

\bibitem[{\citenamefont{Will}(2014)}]{Will_2014}
\bibinfo{author}{\bibfnamefont{C.~M.} \bibnamefont{Will}},
  \bibinfo{journal}{Living Reviews in Relativity} \textbf{\bibinfo{volume}{17}}
  (\bibinfo{year}{2014}), ISSN \bibinfo{issn}{1433-8351},
  \urlprefix\url{http://dx.doi.org/10.12942/lrr-2014-4}.

\bibitem[{\citenamefont{Nakamura and Koyama}(1982)}]{Nakamura92}
\bibinfo{author}{\bibfnamefont{Y.}~\bibnamefont{Nakamura}} \bibnamefont{and}
  \bibinfo{author}{\bibfnamefont{J.}~\bibnamefont{Koyama}},
  \bibinfo{journal}{Journal of Geophysical Research: Solid Earth}
  \textbf{\bibinfo{volume}{87}}, \bibinfo{pages}{4855} (\bibinfo{year}{1982}),
  \urlprefix\url{https://agupubs.onlinelibrary.wiley.com/doi/abs/10.1029/JB087iB06p04855}.

\end{thebibliography}

\end{document}